\documentclass[11pt,a4paper]{article}
\pdfoutput=1
\usepackage{jheppub}
\usepackage{amsmath}
\usepackage{epsfig}
\usepackage{amssymb}
\usepackage{graphics}
\usepackage[active]{srcltx}

\newcommand{\bea}{\begin{eqnarray}}
\newcommand{\eea}{\end{eqnarray}}
\newcommand{\bean}{\begin{eqnarray*}}
\newcommand{\eean}{\end{eqnarray*}}

\def\WH #1{\widehat{#1}}

\def\eref#1{(\ref{#1})}

\def\a{{\alpha}}

\def\b{{\beta}}

\newcommand{\cA}{{\cal A}}

\newcommand{\cF}{{\cal F}}
\newcommand{\cR}{{\cal R}}

\newcommand{\cI}{{\cal I}}
\newcommand{\cQ}{{\cal Q}}
\newcommand{\cT}{{\cal T}}

\def\Label#1{\label{#1}%
  \smash{\hbox to0pt{\raise1ex\hbox{\tiny[#1]}\hss}}}

\def\oneloop{\tiny\mbox{1-loop}}

\def\tree{\tiny\mbox{tree}}
\def\cyclic#1{\mbox{Cyclic}\{#1\}}

\def\spaa #1{\langle #1\rangle}
\def\spbb #1{[#1]}
\def\spab #1{\langle #1]}

\title{Note on recursion relations for the $\mathcal{Q}$-cut representation}
\author[a,b]{Bo Feng,}
\author[c,d]{Song He,}
\author[a]{Rijun Huang\footnote{The correspondence author.}}
\author[a]{and Ming-xing Luo}
\affiliation[a]{Zhejiang Institute of Modern Physics, Department of Physics,
 Zhejiang University,\\
 No.38, Zheda Road, Hangzhou, 310027, P.R. China.}
\affiliation[b]{Center of Mathematical Science,
  Zhejiang University,\\
  No.38, Zheda Road, Hangzhou, 310027, P.R. China.}
\affiliation[c]{CAS Key Laboratory of Theoretical Physics, Institute of Theoretical Physics,
Chinese Academy of Sciences,\\
No. 55, ZhongGuanCun East Street, Beijing 100190, P.R.China.}
\affiliation[d]{School of Physical Sciences, University of Chinese Academy of Sciences,\\
No. 19A, Yuquan Road, Beijing 100049, P.R.China.}

\emailAdd{fengbo@zju.edu.cn}
\emailAdd{songhe@itp.ac.cn}
\emailAdd{huang@nbi.dk}
\emailAdd{mingxingluo@zju.edu.cn}

\date{\today}
\abstract{In this note, we study the $\mathcal{Q}$-cut representation by combining it with BCFW deformation.
As a consequence, the one-loop integrand is expressed in terms of a recursion relation, {\it i.e.}, $n$-point one-loop integrand is constructed using tree-level amplitudes and $m$-point one-loop integrands with $m\leq n-1$. By giving explicit examples, we show that the integrand from the recursion relation is equivalent to
that from Feynman diagrams or the original $\mathcal{Q}$-cut construction, up to scale free terms.   }
\keywords{Scattering Amplitude, Loop Integrand}

\begin{document}
\maketitle \flushbottom

%\newpage

%%%%%%%%%%%%%%%%%%%%
\section{Introduction}
%%%%%%%%%%%%%%%%%%%%

In a recent work, a new representation of the perturbative
$S$-matrix, known as $\mathcal{Q}$-cut representation, was proposed
\cite{Baadsgaard:2015twa}. It allows one to write the integrand
of loop amplitude as summation of products of lower-point tree-level
amplitudes with deformed loop momenta. For generic $n$-point
one-loop integrand with all massless external legs, the new
representation takes the form,
\bea
\mathcal{I}^{\mathcal{Q}}_n(\ell)=\sum_{P_L}\sum_{h_1,h_2}\mathcal{A}_L(\cdots,\widehat{\ell}_R^{~h_1},-\widehat{\ell}_L^{~h_2}){1\over
\ell^2(-2\ell\cdot
P_L+P_L^2)}\mathcal{A}(\widehat{\ell}_L^{~\bar{h}_2},-\widehat{\ell}_R^{~\bar{h}_1},\cdots)~,~~~\eea
where $\widehat{\ell}=\alpha_L(\ell+\eta)$, $\widehat{\ell}_R\equiv
\widehat{\ell}_L-P_L$ with $\alpha_L=P_L^2/(2\ell\cdot P_L)\neq 0$,
$\eta^2=\ell^2$. As will be reviewed shortly, two deformations have been applied to the loop
momentum $\ell$: firstly the dimensional deformation $\ell \to
\ell+\eta$ with $\eta$ in extra dimensions, and secondly the scale deformation $\ell\to \alpha\ell$.
The details of the one-loop $\mathcal{Q}$-cut construction was
further clarified in \cite{Huang:2015cwh}, and generalizations to two
loops or more was also illustrated in \cite{Baadsgaard:2015twa}. The
$\mathcal{Q}$-cut representation circumvented two difficulties in the
attempt for recursive construction of loop integrand: canonical
definition of loop momentum and the singularities in the forward limit (which will be referred to as forward singularities). On the other hand, the integration over loop momentum with such integrand still requires more systematic investigations.

The $\mathcal{Q}$-cut representation was partly inspired by the work \cite{Geyer:2015bja}
and finds direct application
% one-loop $\mathcal{Q}$-cut representation is reported to be exist during
in the study of writing one-loop amplitudes based on the Riemann sphere
\cite{Geyer:2015jch, Baadsgaard:2015hia, He:2015yua}\footnote{In the
scattering equation formalism
\cite{Cachazo:2013gna,Cachazo:2013hca,Cachazo:2013iea,Cachazo:2014nsa,Cachazo:2014xea},
loop integrands for super-gravity and super-Yang-Mills amplitude has
formerly been proposed \cite{Geyer:2015bja}, since in these theories
there is no forward singularity.}, and very recently in an extension to two-loop supersymmetric amplitudes from Riemann sphere~\cite{Geyer:2016wjx}.
Another work also reports similar one-loop
integrand expansion while investigating elliptic scattering
equations at one-loop level \cite{Cardona:2016bpi}, based on an earlier work on the $\Lambda$ scattering equation
\cite{Gomez:2016bmv}. The idea in the $\mathcal{Q}$-cut construction
also inspires some thoughts in the other approach of constructing
one-loop amplitude \cite{Cachazo:2015aol}, as well as the
construction of two-loop planar integrand of cubic scalar theory
\cite{Feng:2016nrf}. These works have shown the universality and importance
of $\mathcal{Q}$-cut representation for loop integrands in general.

After the discovery of Britto-Cachazo-Feng-Witten(BCFW) recursion relations for tree-level
amplitudes~\cite{Britto:2004ap,Britto:2005fq}, it is very natural to ask if one can construct loop integrands in a similar, recursive way.  The key for the progress lies in expressing planar loop integrands from forward limits of tree amplitudes~\cite{CaronHuot:2010zt,ArkaniHamed:2010kv,Boels:2010nw}, which has been very successful for cases without forward singularities, such as super-Yang-Mills at one loop and planar ${\cal N}=4$ SYM to all loops~\cite{ArkaniHamed:2010kv}. However, for general theories the afore-mentioned difficulties have only been resolved in the $\mathcal{Q}$-cut construction. These works have indicated clearly  that for generic
loop integrands, BCFW deformation has to be applied with extra care, especially due to the presence of forward
singularities. %\footnote{Recent discussions on the loop integrand representation from forward limit can be found in \cite{Baadsgaard:2015hia,He:2015yua}}.
In the $\mathcal{Q}$-cut construction, the dimensional deformation transforms one-loop
integrand into tree diagrams,
%as well as avoids the forward singularity.
while the scale deformation has avoided the forward singularities by excluding the tree diagrams
that corresponding to one-loop tadpole and massless bubble
contributions, which should not be presented in the final amplitude.

Both recursion relations and $\mathcal{Q}$-cut approach to the
construction of loop integrands in general theories are promising
but with some unsatisfying features: the $\mathcal{Q}$-cut
representation has non-standard propagators, while it is not clear
how to remove forward singularities in general in recursion
relations. Thus it is natural to see if by combining the two methods
to make further progress. In this note, we will initiate the study
along this direction. We would like to see if there is another way
to deal with forward singularities and how much can we learn about
the structure of one-loop integrands from both recursion and
$\mathcal{Q}$-cut viewpoints.

%This motives us to think two questions: is scale deformation the only possible way leading to the final result, and can we apply the BCFW deformation after the dimensional deformation but before the scale deformation? In the following sections, we shall clarify this confusion by inspecting the BCFW deformation in $\mathcal{Q}$-cut construction at one-loop level.

This paper is structured as follows. In \S \ref{secDerivation}, we
illustrate the application of BCFW deformation in the $\mathcal{Q}$-cut construction, and present a recursive formula for
one-loop integrand. In \S \ref{secExample}, we explain the details
of the recursive formula by three examples, and confirm the
validity of the results by comparing with results from one-loop Feynman diagrams and those from the $\mathcal{Q}$-cut construction. We conclude in \S
\ref{secConclusion}.

%%%%%%%%%%%%%%%%%
\section{The derivation of recursion relation}
\label{secDerivation}
%%%%%%%%%%%%%%%%%%%

Let us first recall the original derivation of $\mathcal{Q}$-cut
representation in~\cite{Baadsgaard:2015twa}. After imposing the
dimensional deformation $\ell\to\ell+\eta$  as well
as the shift $\ell\to \ell+P$ for loop momentum, the $n$-point
one-loop integrand $\mathcal{I}^{\mathcal{Q}}({\ell})$ becomes
essentially the $(n+2)$-point tree-level amplitude $\mathcal{T}(\ell)$,
on the condition $\ell^2=0$. Then by scale deformation $\ell\to
\alpha\ell$, and by removing diagrams that contribute to
one-loop tadpoles and massless bubbles appropriately, one gets the one-loop
integrand. Since BCFW recursion has been
applied to the computation of ordinary tree-level amplitudes, this naturally motivates us to consider the possibility of
constructing the $(n+2)$-point tree-level amplitude$\mathcal{T}(\ell)$ using the recursion. Here we present a derivation of the recursive representation for one-loop integrand
following the afore-mentioned motivation. The derivation will take three steps, as follows.

%%%%%%%%%%%%%%%%%%%%%%%%%
\subsection{Step one: dimensional deformation}
%%%%%%%%%%%%%%%%%

Just like the original $\mathcal{Q}$-cut construction
\cite{Baadsgaard:2015twa}, the first step of the derivation is to
reformulate one-loop integrand in terms of tree-level amplitudes. We take
the same dimensional deformation $\ell\to \ell+\eta$ as in
\cite{Baadsgaard:2015twa} and also the loop momentum shifting, to
arrive at
\bea \mathcal{A}^{\oneloop}=\int d^{D}\ell~
\mathcal{I}^{\mathcal{Q}}(\ell)~~~,~~~\mathcal{I}^{\mathcal{Q}}(\ell)={1\over
\ell^2}\mathcal{T}^{\mathcal{Q}}(\ell)~.~~~\label{recurStep1}\eea
Some explanations are in order for (\ref{recurStep1}).
Firstly, from the dimensional deformation, it is known that
$\mathcal{T}^{\mathcal{Q}}$ is given by those Feynman diagrams with
$n$ external legs and two extra legs by cutting an internal
propagator. Thus $\mathcal{T}^{\mathcal{Q}}$ is defined on the
condition $\ell^2=0$, which says that all $\ell$ in
$\mathcal{T}^{\mathcal{Q}}$ should be understood as the null
momentum in higher dimension. Furthermore,
$\mathcal{T}^{\mathcal{Q}}$ is not exactly the full $(n+2)$-point
tree-level amplitude, since in order to reconstruct the one-loop
integrand, some diagrams should be excluded. Such tree-level
diagrams correspond to one-loop tadpole and massless bubble diagrams
with single cuts. From Feynman diagrams one can inspect that, a
tadpole after single cut will produce tree diagrams with
$\ell,-\ell$ attaching to the same vertex\footnote{Here $\ell,-\ell$
denotes two legs by breaking an internal line.}, while massless
bubble diagram with the massless leg $p_i$ after single cut will
produce tree diagrams with $\ell, p_i$ (or $-\ell, p_i$) attaching
to the same three-point vertex, and then meeting $-\ell$ (or $\ell$)
in the neighboring vertex. The above scenery would help us to
exclude corresponding tree diagrams in the following steps.

Next let us take a look at the contributing tree diagrams to
$\mathcal{T}^{\mathcal{Q}}$. If the theory under consideration is
not color-ordered, we shall consider the full $(n+2)$-point on-shell
tree-level Feynman diagrams after removing those corresponding to
the one-loop tadpole and massless bubbles. While if it is
color-ordered, the $\mathcal{T}^{\mathcal{Q}}$ gets contribution
from $n$ different color-ordered tree diagrams, each by breaking an internal
line of the $n$ propagators. Since there are $n$ different color orderings,
we can calculate each one independently, for example, using different methods
(such as Feynman diagrams or BCFW recursion relations) or different deformations
in BCFW recursion relations.

A final remark says that, the loop momentum shifting in expression
(\ref{recurStep1}) makes a canonical definition of loop momentum,
such that the integrand is irrelevant to the labeling of $\ell$ for
internal propagators.

%%%%%%%%%%%%%%%%%%%%
\subsection{Step two: BCFW deformation}
%%%%%%%%%%%%%%%%%%%%%

Now let us turn to $\mathcal{T}^{\mathcal{Q}}$, and our aim is to
determine it by BCFW deformation. Since it is effectively tree-level
amplitude but with forward singularity removed, the analysis on the
large $z$ behavior would be the same and the computation should be
straightforward. Let us, for generality, take two arbitrary momenta
$p_i,p_j$ (but not $\ell,-\ell$) and perform the standard BCFW
deformation
\bea
\widehat{p}_i=p_i+zq~~~,~~~\widehat{p}_j=p_j-zq~~\mbox{with}~~q^2=q\cdot
p_i=q\cdot p_j=0~.~~~\eea
Such deformation can be realized when the dimension $D\geq 4$. In
this case, $\mathcal{T}^{\mathcal{Q}}$ becomes an analytic function
of external momenta $p_i$'s, loop momentum $\ell$ and a complex
variable $z$. As usual, we can consider the contour integration
\bea \oint_{\Gamma} {dz\over
z}~\mathcal{T}^{\mathcal{Q}}(z)~,~~~\eea
where the contour $\Gamma$ is a very large circle. This integration
leads to
\bea \mathcal{T}^{\mathcal{Q}}(z=0)=\mathcal{B}+\sum_{z=z_\gamma}
{\mathcal{T}^{\mathcal{Q}}\over z}~,~~~\label{I-gen-2}\eea
where the sum is over all finite pole $z_\gamma$'s of
$\mathcal{T}^{\mathcal{Q}}$, and $\mathcal{B}$ is possible boundary
contribution. It is well-known for tree-level amplitudes that for
Yang-Mills and gravity theories, the BCFW deformation can be chosen
such that the boundary contribution vanishes. While for some other
theories, the boundary contribution would appear and require more
careful analysis
\cite{ArkaniHamed:2008yf,Cheung:2008dn,Feng:2009ei,Jin:2014qya,Jin:2015pua,Feng:2014pia,Feng:2015qna,Cheung:2015cba,Cheung:2015ota}.
Here we shall assume $\mathcal{B}=0$ for simplicity (but the similar
consideration can be generalized to the case with non-zero boundary
contributions). Thus the only information we need for computing
$\mathcal{T}^{\mathcal{Q}}$ by means of expression (\ref{I-gen-2})
is the pole structure of function $\mathcal{T}^{\mathcal{Q}}(z)$.

The BCFW deformation splits a tree amplitude into two parts, with
the shifted momenta $\widehat{p}_i,\widehat{p}_j$ locating in each
part. Assuming $\widehat{K}_\gamma\equiv\widehat{p_i}+P_\gamma$ is
the sum of all momenta in the part containing $\widehat{p}_i$, and
$K_\gamma\equiv p_i+P_\gamma$. From $\widehat{K}_{\gamma}^2=0$ we
get $z_\gamma=-K_\gamma^2/(2q\cdot K_\gamma)$. Now let us consider
the two extra legs $\ell,-\ell$. If they are in the same part,
$K_\gamma$ will have no dependence on $\ell$, thus also the pole
$z_\gamma$. We shall denote the corresponding contribution as
$\mathcal{R}_A^{\mathcal{Q}}$. While if $\ell, -\ell$ are separated
in two parts, $K_\gamma$ as well as $z_\gamma$ would depend on
$\ell$. We shall denote the corresponding contribution as
$\mathcal{R}_{B}^{\mathcal{Q}}$. So we have
\bea
\mathcal{T}^{\mathcal{Q}}=\mathcal{R}^{\mathcal{Q}}_A+\mathcal{R}^{\mathcal{Q}}_B~.~~~\label{I-gen-2-1}\eea

For the contribution $\mathcal{R}^{\mathcal{Q}}_A$, we can further
organize it into two parts,
\bea
\mathcal{R}^{\mathcal{Q}}_A=\mathcal{R}^{\mathcal{Q}}_{A,1}+\mathcal{R}^{\mathcal{Q}}_{A,2}~.~~~\eea
$\mathcal{R}^{\mathcal{Q}}_{A,1}$ denotes the contribution where
legs $\ell,-\ell$ are in the part containing $\widehat{p_j}$, while
$\mathcal{R}^{\mathcal{Q}}_{A,2}$ denotes the contribution where
legs $\ell,-\ell$ are in the part containing $\widehat{p_i}$.
Explicitly, we have
\bea
\mathcal{R}^{\mathcal{Q}}_{A,1}=\sum_{h,\gamma}A(\widehat{p}_i(z_\gamma),\{\gamma\},-\widehat{K}_\gamma^{h}(z_\gamma)){1\over
K_\gamma^2}\mathcal{T}(\widehat{K}_\gamma^{-h}(z_\gamma),\widehat{p}_j(z_\gamma),
\{\beta\},\ell,-\ell)~,~~~\label{recurStep2A1}\eea
where
\bea z_\gamma=-{(P_\gamma+p_i)^2\over 2q\cdot
P_\gamma}~~~,~~~\widehat{K}_\gamma(z_\gamma)=P_\gamma+p_i+z_\gamma
q~,~~~\nonumber\eea
as well as $\widehat{p}_i(z_\gamma)=p_i+z_\gamma q$,
$\widehat{p}_j(z_\gamma)=p_j-z_\gamma q$, and $\{\gamma\}\cup
\{\b\}=\{1,2,\ldots, n\}/\{i,j\}$. Similarly,
\bea \mathcal{R}^{\mathcal{Q}}_{A,2}=\sum_{h,
\beta}\mathcal{T}(\ell,-\ell,\{\gamma\},\widehat{p_i}(z_\b),-\widehat{K}^{h}_\b(z_\b)){1\over
K_\b^2}A(\widehat{K}_\b^{-h}(z_\b),\{\beta\},\widehat{p}_j(z_\b))~,~~~\label{recurStep2A2}\eea
where
\bea z_\b={(P_\b+p_j)^2\over 2q\cdot
P_\b}~~~,~~~\widehat{K}_\b(z_\b)=-(P_\b+p_j-z_\b
q)~.~~~\nonumber\eea
Note that the sum is over all possible splitting of $(n-2)$ legs
$\{1,2,\ldots, n\}/\{i,j\}$ and helicities. Also note that inside
the bracket $A(\bullet), \mathcal{T}(\bullet)$ we have explicitly
labeled all the legs in each part but not the ordering of legs. The
color-ordering of legs should be understood with respect to their
corresponding theories.

Now let us take a more careful look on expressions
(\ref{recurStep2A1}) and (\ref{recurStep2A2}). Firstly, the
$\mathcal{T}$ part in $\mathcal{R}^{\mathcal{Q}}_{A,1},
\mathcal{R}^{\mathcal{Q}}_{A,2}$ will be lower-point on-shell tree
diagrams after excluding those corresponding to tadpole and bubble
diagrams. This means that when dressing with ${1\over \ell^2}$, they
would become lower-point one-loop integrand, which can be obtained
by any legitimate methods, such as the original $\mathcal{Q}$-cut construction or Feynman diagram
method with partial fraction identity. One important implication
 is that the forward singularities in the type $\mathcal{R}_A$
have been automatically removed after using the well-defined
one-loop integrands of lower points. Secondly, for
$\mathcal{R}^{\mathcal{Q}}_{A,1}$, the number of legs in set
$\{\gamma\}$ must be at least one, in order for the amplitude to be
non-vanishing. Naively, the number of legs in set $\{\beta\}$ could
also be zero. However, when it is so, the tree diagrams of
$\mathcal{T}$ are exactly those corresponding to tadpole and
massless bubbles, which need to be excluded. So $\{\beta\}$ could
not be empty set. Similarly for $\mathcal{R}^{\mathcal{Q}}_{A,2}$,
the number of legs in sets $\{\gamma\}, \{\b\}$ should at least be
one.

Now let us analyze the contribution $\mathcal{R}^{\mathcal{Q}}_B$.
We can also organize it into two parts,
\bea
\mathcal{R}^{\mathcal{Q}}_{B}=\mathcal{R}^{\mathcal{Q}}_{B,1}+\mathcal{R}^{\mathcal{Q}}_{B,2}~.~~~\eea
$\mathcal{R}^{\mathcal{Q}}_{B,1}$ denotes the contribution where leg
$\ell$ is in the part containing $\widehat{p}_i$, while
$\mathcal{R}^{\mathcal{Q}}_{B,2}$ denotes the contribution where leg
$\ell$ is in the part containing $\widehat{p}_j$, explicitly as
\bea
\mathcal{R}^{\mathcal{Q}}_{B,1}=\sum_{h,\gamma}\mathcal{T}(\ell,\widehat{p}_i(z_\gamma),\{\gamma\},-\widehat{K}^{h}_\gamma(z_\gamma)){1\over
K_\gamma^2}\mathcal{T}(\widehat{K}^{-h}_\gamma(z_\gamma),\widehat{p}_j(z_\gamma),\{\b\},-\ell)~,~~~\label{recurStep2B1}\eea
where
\bea z_\gamma=-{(P_\gamma+p_i+\ell)^2\over 2 q \cdot
(P_\gamma+\ell)}~~~,~~~\widehat{K}_\gamma(z_\gamma)=P_\gamma+p_i+\ell+z_\gamma
q~,~~~\nonumber\eea
and $\{\gamma\}\cup \{\b\}=\{1,2,\ldots, n\}/\{i,j\}$. While
\bea
\mathcal{R}^{\mathcal{Q}}_{B,2}=\sum_{h,\gamma}\mathcal{T}(-\ell,\widehat{p}_i(z_\gamma),\{\gamma\},-\widehat{K}^{h}_\gamma(z_\gamma)){1\over
K_\gamma^2}\mathcal{T}(\widehat{K}^{-h}_\gamma(z_\gamma),\widehat{p}_j(z_\gamma),\{\b\},\ell)~,~~~\label{recurStep2B2}\eea
where
\bea z_\gamma=-{(P_\gamma+p_i-\ell)^2\over 2 q \cdot
(P_\gamma-\ell)}~~~,~~~\widehat{K}_\gamma(z_\gamma)=P_\gamma+p_i-\ell+z_\gamma
q~.~~~\nonumber\eea
Some discussions are in order for expressions (\ref{recurStep2B1})
and (\ref{recurStep2B2}). Notice that we have used $\mathcal{T}$
instead of tree-level amplitude $A$, since in this stage potential
contributions coming from  corresponding to tadpole and bubble
diagrams in $\mathcal{R}^{\mathcal{Q}}_{B,1},
\mathcal{R}^{\mathcal{Q}}_{B,2}$ should be excluded. Recalling our
discussion on the excluded diagrams in the previous subsection, we
can conclude that, since $\ell,-\ell$ are separated into two parts,
there could not be diagrams corresponding to one-loop tadpoles,
while diagrams corresponding to massless bubbles\footnote{We need to
distinguish massless bubble from massive bubble. The latter is
allowed for one-loop diagrams.} do exist in
$\mathcal{R}^{\mathcal{Q}}_{B,1}$ and
$\mathcal{R}^{\mathcal{Q}}_{B,2}$ when the set $\{\gamma\}$ or
$\{\beta\}$ is empty. In other words, forward singularities
corresponding to tadpoles have been avoided in type $\mathcal{R}_B$.
Combining the discussions for type $\mathcal{R}_A$, we see that we
can remove forward singularities corresponding to tadpoles without
using scale deformation as is done in the $\mathcal{Q}$-cut
construction. However, forward singularities that corresponding to
massless bubbles are more difficult to deal with and we will
organize $\mathcal{R}^{\mathcal{Q}}_{B,1}$ into three contributions
\bea
\mathcal{R}^{\mathcal{Q}}_{B,1}=\mathcal{R}'_{B,1}+\mathcal{R}''_{B,1}+\mathcal{R}'''_{B,1}~.~~~\label{recurStep2B3}\eea
$\mathcal{R}'_{B,1}$ denotes the contribution of the case when both
$\{\gamma\}$ and $\{\b\}$ are not empty, so  forward singularities corresponding
to massless bubbles will not appear and  there will be no
excluded diagrams. Thus the $\mathcal{T}$ is exactly the tree
amplitude and we  have
\bea \mathcal{R}'_{B,1}=\sum_{\gamma, h}^{1\leq |\gamma|\leq n-3}
A(\ell,\widehat{p}_i(z_\gamma),\{\gamma\},-\widehat{K}^{h}_{\gamma}(z_\gamma)){1\over
K_\gamma^2}A(\widehat{K}^{-h}_{\gamma}(z_\gamma),\widehat{p}_j(z_\gamma),\{\b\},-\ell)~,~~~\label{recurStep2B31}\eea
where the sum is over all helicities and possible splitting of
external legs with the length of set $\{\gamma\}$ satisfying $1\leq
|\gamma| \leq n-3$. This is to ensure that there is at least one leg
in set $\{\gamma\}, \{\b\}$.

$\mathcal{R}''_{B,1}$ denotes the special case when set
$\{\gamma\}=\emptyset$. In this case, $\mathcal{T}(\ell,
\widehat{p}_i,\{\gamma\},-\widehat{K}_{\gamma})$ becomes a
three-point amplitude, and we get explicitly
\bea
\mathcal{R}''_{B,1}=\sum_{h}A(\ell,\widehat{p}_i(z_\gamma),-\widehat{K}^{h}_{\gamma}(z_\gamma)){1\over
2\ell\cdot p_i}
\mathcal{T}(\widehat{K}^{-h}_{\gamma}(z_\gamma),\widehat{p}_j(z_\gamma),\{\b\},-\ell)~,~~~\label{recurStep2B32}\eea
where
\bea z_\gamma=-{2p_i \cdot \ell\over 2q\cdot
\ell}~~~,~~~\widehat{K}_\gamma(z_\gamma)=\ell+p_i+z_\gamma
q~,~~~\nonumber\eea
and $\{\b\}=\{1,2,\ldots, n\}/\{i,j\}$.

$\mathcal{R}'''_{B,1}$ denotes the special case when set
$\{\b\}=\emptyset$. In this case,
$\mathcal{T}(\widehat{K}_\gamma,\widehat{p}_j,\{\b\},-\ell)$ becomes
a three-point amplitude, and we get explicitly
\bea
\mathcal{R}'''_{B,1}=\sum_{h}\mathcal{T}(\ell,\widehat{p}_i(z_\gamma),\{\gamma\},-\widehat{K}^{h}_{\gamma}(z_\gamma)){1\over
-2\ell\cdot
p_j}A(\widehat{K}^{-h}_\gamma(z_\gamma),\widehat{p}_j(z_\gamma),-\ell)~,~~~\label{recurStep2B33}\eea
where
\bea z_\gamma={2p_j\cdot \ell\over 2q\cdot
\ell}~~~,~~~\widehat{K}_\gamma(z_\gamma)=-(p_j-\ell-z_\gamma
q)~,~~~\eea
and $\{\gamma\}=\{1,2,\ldots,n\}/\{i,j\}$.

Similarly, we can also organize $\mathcal{R}^{\mathcal{Q}}_{B,2}$
into three parts,
\bea
\mathcal{R}^{\mathcal{Q}}_{B,2}=\mathcal{R}'_{B,2}+\mathcal{R}''_{B,2}+\mathcal{R}'''_{B,2}~,~~~\label{recurStep2B4}\eea
just as it is defined for $\mathcal{R}^{\mathcal{Q}}_{B,1}$, but
changing $\ell\to -\ell$. Explicitly, we have
\bea \mathcal{R}'_{B,2}=\mathcal{R}'_{B,1}|_{\ell\to
-\ell}~,~~~\label{recurStep2B4}\eea
and $\mathcal{R}''_{B,2}=\mathcal{R}''_{B,1}|_{\ell\to-\ell}$,
$\mathcal{R}'''_{B,2}=\mathcal{R}'''_{B,1}|_{\ell\to-\ell}$.

There is an important observation. If we consider the color-ordered
integrand, we can choose the deformation pair $(i,j)$ such that
$\ell, -\ell$ are not nearly with the deformed momenta. Thus the
contributions of $\mathcal{R}''_{B,2}$, $\mathcal{R}''_{B,1}$,
$\mathcal{R}'''_{B,2}$ and $\mathcal{R}'''_{B,1}$ do not exist. As
we will discuss in the following subsection, the remaining forward
singularities that corresponding to massless bubbles are exactly in
those four terms. In other words, with a proper choice of
deformation pair, we can naturally avoid forward singularities
without further using the scale deformation.

%%%%%%%%%%%%%%%%%%%%
\subsection{Step three: scale deformation}
%%%%%%%%%%%%%%%%

In the previous subsection we have expressed
$\mathcal{T}^{\mathcal{Q}}$ as
\bea
\mathcal{T}^{\mathcal{Q}}=\mathcal{R}^{\mathcal{Q}}_{A}+\mathcal{R}^{\mathcal{Q}}_{B}~,~~~\eea
where
$\mathcal{R}^{\mathcal{Q}}_{A}=\mathcal{R}^{\mathcal{Q}}_{A,1}+\mathcal{R}^{\mathcal{Q}}_{A,2}$
given in expressions (\ref{recurStep2A1}), (\ref{recurStep2A2})
respectively, and
$\mathcal{R}^{\mathcal{Q}}_{B}=\mathcal{R}^{\mathcal{Q}}_{B,1}+\mathcal{R}^{\mathcal{Q}}_{B,2}$,
with
$\mathcal{R}^{\mathcal{Q}}_{B,1}=\mathcal{R}'_{B,1}+\mathcal{R}''_{B,1}+\mathcal{R}'''_{B,1}$
given in expressions (\ref{recurStep2B31}), (\ref{recurStep2B32}),
(\ref{recurStep2B33}), and
$\mathcal{R}^{\mathcal{Q}}_{B,2}=\mathcal{R}'_{B,2}+\mathcal{R}''_{B,2}+\mathcal{R}'''_{B,2}$
by changing $\ell\to -\ell$ of $\mathcal{R}^{\mathcal{Q}}_{B,1}$. In
each $\mathcal{R}$ expression there would be $\mathcal{T}$
functions, and we should identify them. The $\mathcal{T}$ functions
are determined by removing  tree diagrams that corresponding to
tadpole and massless bubbles. In the previous subsections, we have
presented some discussions on this point, but the complete
resolution will be provided in this subsection. In fact, as we have
pointed out, the only left forward singularities are those in terms
$\mathcal{R}''_{B,1},\mathcal{R}'''_{B,1}$ and
$\mathcal{R}''_{B,2},\mathcal{R}'''_{B,2}$. To deal with them, we
use the scale deformation.
%the only terms that need to be justified by scale deformation are the special cases $\mathcal{R}''_{B,1},\mathcal{R}'''_{B,1}$ and $\mathcal{R}''_{B,2},\mathcal{R}'''_{B,2}$, which will be treated

Before giving a careful discussion,  let us take a look on
$\mathcal{R}^{\mathcal{Q}}_{A,1}$,
$\mathcal{R}^{\mathcal{Q}}_{A,2}$. When multiplying ${1\over
\ell^2}$ with $\mathcal{T}$ in (\ref{recurStep2A1}),
(\ref{recurStep2A2}), it trivially becomes one-loop integrand of the
original $\mathcal{Q}$-cut representation with BCFW-deformed
momenta. Thus we can identify them as
\bea {\mathcal{R}^{\mathcal{Q}}_{A,1}\over
\ell^2}=\sum_{h,\gamma}A(\widehat{p}_i(z_\gamma),\{\gamma\},-\widehat{K}_\gamma^{h}(z_\gamma)){1\over
K_\gamma^2}\mathcal{I}^{\mathcal{Q}}(\widehat{K}_\gamma^{-h}(z_\gamma),\widehat{p}_j(z_\gamma),
\{\beta\},\ell,-\ell)~,~~~\label{recurStep3A1}\eea
where $z_\gamma=-{(P_\gamma+p_i)^2\over 2q\cdot P_\gamma}$,
$\widehat{K}_\gamma(z_\gamma)=P_\gamma+p_i+z_\gamma q$. Similarly,
\bea {\mathcal{R}^{\mathcal{Q}}_{A,2}\over \ell^2}=\sum_{h,
\beta}\mathcal{I}^{\mathcal{Q}}(\ell,-\ell,\{\gamma\},\widehat{p_i}(z_\b),-\widehat{K}^{h}_\b(z_\b)){1\over
K_\b^2}A(\widehat{K}_\b^{-h}(z_\b),\{\beta\},\widehat{p}_j(z_\b))~,~~~\label{recurStep3A2}\eea
where $z_\b={(P_\b+p_j)^2\over 2q\cdot P_\b}$,
$\widehat{K}_\b(z_\b)=-(P_\b+p_j+z_\b q)$. Here
$\mathcal{I}^{\mathcal{Q}}$'s are lower-point one-loop integrands
from $\mathcal{Q}$-cut representation, and $A$'s are lower-point
tree amplitudes. In fact, the one-loop integrand in \eref{recurStep3A1} and \eref{recurStep3A2}
does not need to be in $\mathcal{Q}$-cut representation, i.e., any representation, such as
the one obtained by Feynman diagrams, should be fine.
Thus these two terms can be expressed as summation
over products of lower-point one-loop integrand and tree amplitude.
For other two terms $\mathcal{R}'_{B,1}, \mathcal{R}'_{B,2}$, it has
already been shown in (\ref{recurStep2B31}) that they are summation
over products of two lower-point tree amplitudes. The important point is that
for these two terms, the loop momentum $\ell$ is not scaled.

Now let us focus on the special cases
$\mathcal{R}''_{B,1},\mathcal{R}'''_{B,1}$,
$\mathcal{R}''_{B,2},\mathcal{R}'''_{B,2}$, and specifically take
$\mathcal{R}''_{B,1}$
\bea
\mathcal{R}''_{B,1}=\sum_{h}A(\ell,\widehat{p}_i(z_\gamma),-\widehat{K}^{h}_{\gamma}(z_\gamma)){1\over
2\ell\cdot p_i}
\mathcal{T}(\widehat{K}^{-h}_{\gamma}(z_\gamma),\widehat{p}_j(z_\gamma),\{1,\ldots,n\}/\{i,j\},-\ell)~~~~\eea
as example. We need to exclude the contribution of massless bubbles
from it. In order to do so, let us introduce a scale deformation
$\ell\to \alpha\ell$ as is done in the original $\mathcal{Q}$-cut
construction. Since $z_\gamma=-{2p_i\cdot \ell\over 2 q\cdot \ell}$,
the scale deformation will not change the location of pole
$z_{\gamma}$. Hence we can write $\mathcal{R}''_{B,1}$ as
\bea
\mathcal{R}''_{B,1}(\alpha)=\sum_{h}A(\alpha\ell,\widehat{p}_i(z_\gamma),-\widehat{K}^{h}_{\gamma}(z_\gamma,\alpha)){1\over
2\ell\cdot p_i}
\mathcal{T}(\widehat{K}^{-h}_{\gamma}(z_\gamma,\alpha),\widehat{p}_j(z_\gamma),\{1,\ldots,n\}/\{i,j\},-\alpha\ell)~,~~~\label{recurStep3B}\eea
where
$\widehat{K}_{\gamma}(z_{\gamma},\alpha)=\alpha\ell+p_i+z_{\gamma}q$.

Let us have a more detailed discussion on the
$\mathcal{T}(\widehat{K}_\gamma,\widehat{p}_j,\{1,\ldots,n\}/\{i,j\},-\alpha\ell)$
of (\ref{recurStep3B}). The on-shell condition of
$\widehat{K}_{\gamma}$ is manifestly satisfied for any value of $\a$, since (remembering that $q\cdot p_i=0$)
\bea \widehat{K}^2_{\gamma}=(\alpha\ell+p_i-{2p_i\cdot\ell\over
2q\cdot \ell}q)^2=\alpha (2p_i\cdot \ell)-\alpha(2q\cdot
\ell){2p_i\cdot\ell\over 2q\cdot \ell}=0~.~~~\eea
Having verified the on-shell condition, let us concentrate on the
pole structure. We will divide poles into three categories. If the
pole does not contain $-\alpha\ell$ and $\widehat{K}_{\gamma}$, then
it could either be the sum $P$ of some ordinary external legs, or
the one containing $\widehat{p}_j=p_j+{2p_i\cdot\ell\over 2q\cdot
\ell}q$. For the latter case, we have
\bea (P+p_j+{2p_i\cdot\ell\over 2q\cdot \ell}q)^2=(P^2+2P\cdot
p_j)+(2P\cdot q){2p_i\cdot\ell\over 2q\cdot
\ell}={2\big((P^2+2P\cdot p_j)q+(2P\cdot q)p_i\big)\cdot \ell\over
2q\cdot \ell}~.~~~\eea
So this pole is in the scale free form. Similarly, if
$\widehat{p}_j$ appears in the numerator, it will give a
contribution of $q\cdot \ell$ in the denominator. Anyway it is also
in the scale free form. In other words, these poles does not depend
on $\a$ under the scale deformation.

If the pole contains $-\alpha\ell$ or
$\widehat{K}_{\gamma}=\alpha\ell+\widehat{p}_i$, we can always use
momentum conservation to rewrite $\widehat{K}$ as the leg
$-\alpha\ell$, so that the pole is in the form containing
$-\alpha\ell$. For these cases, we can have either
$(P-\alpha\ell)^2=P^2-\a(2P\cdot\ell)$ leading to a finite pole
$\alpha_P={P^2\over 2P\cdot \ell}$, or
\bea (P+p_j-z_\gamma q-\alpha\ell)^2=P^2+2P\cdot p_j+(2P\cdot
q)z_{\gamma}-2\alpha (P+p_j+p_i)\cdot \ell~,~~~\eea
leading to a finite pole
\bea \alpha_P={P^2+2P\cdot p_j+(2P\cdot q)z_{\gamma}\over
2(P+p_i+p_j)\cdot\ell}~.~~~\eea
Note that both solutions depend on the loop momentum $\ell$.

If the pole contains both $-\alpha\ell$ and $\widehat{K}$, then it
has no dependence on $\alpha$. This case contains the contribution
corresponding to massless bubbles which should be excluded. To see
this, let us recall that for the tree diagram that corresponding to
massless bubbles with massless external leg $\widehat{p}_i$, the
legs $\ell, \widehat{p}_i$ are attached to the same three-point
vertex, then they meet leg $-\ell$ in the neighboring vertex.
Explicitly for the tree diagrams of
$\mathcal{T}(\widehat{K},\widehat{p}_j,\{1,\ldots,n\}/\{i,j\},-\ell)$,
it corresponds to the diagrams where legs $\widehat{K}$ and $-\ell$
are attached to the same vertex\footnote{It is easy to see that if
we perform the scale deformation $\ell\to\alpha\ell$, such terms
will not contain $\alpha$ in the denominator.}. This means that the
terms corresponding to the massless bubbles are included in the
boundary part.

Having understood poles of above three categories, we can now consider the following contour
integration
\bea &&\oint {d\alpha\over
\alpha-1}\mathcal{T}(\widehat{K}_{\gamma}(z_\gamma,\alpha),\widehat{p}_j(z_\gamma),\{1,2,\ldots,n\}/\{i,j\},-\alpha\ell)\nonumber\\
&=&\oint {d\alpha\over \alpha-1}{N(-\alpha\ell,\widehat{p}_j)\over
\prod_{\lambda_1}(P_{\lambda_1}+p_j-z_{\gamma}q)^2\prod_{\lambda_2}(P_{\lambda_2}-\alpha\ell)^2\prod_{\lambda_3}(P_{\lambda_3}+p_j-z_{\gamma}q-\alpha\ell)^2}~,~~~\label{contourI2}\eea
where in the second line we have explicitly written down the above
mentioned subtle factors in the denominator. Now we consider its
various pole contributions,
\begin{itemize}
  \item The pole $\alpha=1$ gives the full un-deformed tree
  amplitude.
  \item There are poles at $\alpha=0$. Such poles will appear
  for the propagator $(P_{\lambda_2}-\alpha\ell)^2$ when
  $P_{\lambda_2}^2=0$. The other pole
  $(P_{\lambda_3}+p_j-z_\gamma q-\alpha\ell)^2$ can not
  contribute to $\alpha=0$ pole for generic momentum
  configuration. From expression (\ref{contourI2}) we know that
  the residue at $\a=0$ is scale free term and we can ignore
  them. Note that for this argument to be true, we have assumed
  the factor
  $A(\alpha\ell,\widehat{p}_i(z_{\gamma}),-\widehat{K}_{\gamma}(z_{\gamma},\alpha))$
  in (\ref{recurStep3B}) would not provide denominator that
  breaking the scale free form.
  \item For the pole at $\a=\infty$, it contains the contribution
  from massless bubbles, which should be excluded. However, It
  also contains other contributions which should be included in
  the final result. But inspecting the expression
  (\ref{contourI2}), it can be checked that all such
  contributions are scale free terms, and we can exclude all the
  contributions at $\a=\infty$, letting the result to be valid
  up to some scale free terms.
\end{itemize}
With above consideration, we can claim that, the contributions of
finite $\alpha$ poles are the ones wee need for constructing the
one-loop integrands, without the contributions that corresponding to
tadpole and massless bubbles, and valid up to some scale free terms.
Thus we can write
$\mathcal{T}(\widehat{K}^h_{\gamma}(z_\gamma,\alpha),\widehat{p}_j(z_\gamma),\{1,2,\ldots,n\}/\{i,j\},-\alpha\ell)$
as
\bea \mathcal{T}=\sum_{h',\lambda\in P_{\lambda}^2\neq
0}A(\widehat{K}^h_{\gamma}(z_{\gamma},\alpha_{\lambda}),\{\beta\},K^{-h'}_{\lambda}(\alpha_\lambda)){1\over
P_{\lambda}^2-2P_\lambda\cdot\ell}A(-K^{h'}_{\lambda}(\a_\lambda),\{\lambda\},-\alpha_{\lambda}\ell)~,~~~\eea
where $\alpha_\lambda={P_\lambda^2\over 2P_\lambda\cdot \ell}$,
$K_\lambda(\alpha_{\lambda})=P_\lambda-\alpha_\lambda\ell$,
$\{\beta\}\cup\{\lambda\}=\{1,2,\ldots,n\}/\{i,j\}+\{\widehat{j}\}$,
and the summation is over all possible splitting of
$\{1,2,\ldots,n\}/\{i,j\}+\{\widehat{j}\}$, but with the condition
$P_\lambda^2=0$, which means that the set $\{\lambda\}$ should have
more than one external leg.

With above result, we can finally write the $\mathcal{R}''_{B,1}$ as
\bea
\mathcal{R}''_{B,1}=\sum_{h}A(\ell,\widehat{p_i},-\widehat{K}^{h}_{\gamma}){1\over
2\ell\cdot p_i}\left(\sum_{h',\lambda\in P_\lambda^2\neq
0}A(\widehat{K}^{-h}_{\gamma},\{\beta\},K^{-h'}_{\lambda}){1\over
P_\lambda^2-2P_\lambda\cdot
\ell}A(-K^{h'}_{\lambda},\{\lambda\},-\alpha_\lambda\ell)\right)~,~~~\label{recurStep3B1}\eea
where
\bea z_{\gamma}=-{2p_i\cdot\ell\over 2q\cdot
\ell}~~~,~~~\alpha_\lambda={P_\lambda^2\over
2P_\lambda\cdot\ell}~,~~~\nonumber\eea
and $\widehat{p}_i=p_i+z_{\gamma}q$,
$\widehat{K}_{\gamma}=\alpha_\lambda \ell+p_i+z_\gamma q$,
$K_{\lambda}=P_{\lambda}-\alpha_\lambda\ell$,
$\{\beta\}\cup\{\lambda\}=\{1,2,\ldots,n\}/\{i,j\}+\{\widehat{j}\}$.

Similarly, we have
\bea\mathcal{R}'''_{B,1}=\sum_{h}\left(\sum_{h',\lambda\in
P_\lambda^2\neq
0}A(\alpha_{\lambda}\ell,\{\lambda\},-K_{\lambda}^{h'}){1\over
P_{\lambda}^2+2P_{\lambda}\cdot
\ell}A(K^{-h'}_\lambda,\{\b\},-\widehat{K}^{h}_{\gamma})\right){1\over
-2\ell\cdot
p_j}A(\widehat{K}^{-h}_\gamma,\widehat{p}_j,-\ell)~,~~~\label{recurStep3B2}\eea
where
\bea z_{\gamma}={2p_j\cdot\ell\over 2q\cdot
\ell}~~~,~~~\alpha_\lambda=-{P_{\lambda}^2\over
2P_\lambda\cdot\ell}~,~~~\nonumber\eea
and $\widehat{p}_j=p_j-z_\gamma q$,
$K_\lambda=P_{\lambda}+\alpha_\lambda \ell$,
$K_\gamma=-\a_{\lambda}\ell+p_j-z_{\gamma}q$,
$\{\lambda\}\cup\{\beta\}=\{1,2,\ldots,n\}/\{i,j\}+\{\widehat{i}\}$.

We also have
\bea
\mathcal{R}''_{B,2}=\mathcal{R}''_{B,1}|_{\ell\to-\ell}~~~,~~~\mathcal{R}'''_{B,2}=\mathcal{R}'''_{B,1}|_{\ell\to-\ell}~.~~~\label{recurStep3B3}\eea

To summarize, by BCFW deformation, we have expressed the $n$-point
one-loop integrand recursively as
\bea\mathcal{I}_n={1\over
\ell^2}(\mathcal{R}^{\mathcal{Q}}_A+\mathcal{R}^{\mathcal{Q}}_B)~,~~~\label{recurStepFinal}\eea
where
$\mathcal{R}^{\mathcal{Q}}_A=\mathcal{R}^{\mathcal{Q}}_{A,1}+\mathcal{R}^{\mathcal{Q}}_{A,2}$,
and ${1\over \ell^2}\mathcal{R}^{\mathcal{Q}}_{A,1}$, ${1\over
\ell^2}\mathcal{R}^{\mathcal{Q}}_{A,2}$ are defined as formulas
(\ref{recurStep3A1}), (\ref{recurStep3A2}) respectively, which are
summation of products of lower-point tree amplitude with low-point
one-loop integrand of $\mathcal{Q}$-cut construction. Also,
$\mathcal{R}^{\mathcal{Q}}_{B}=\mathcal{R}'_{B,1}+\mathcal{R}''_{B,1}+\mathcal{R}'''_{B,1}+\mathcal{R}'_{B,2}+\mathcal{R}''_{B,2}+\mathcal{R}'''_{B,2}$.
Among which, $\mathcal{R}'_{B,1}, \mathcal{R}'_{B,2}$ are defined in
formulas (\ref{recurStep2B31}), (\ref{recurStep2B4}) respectively,
which are summation of products of two lower-point tree amplitudes,
and
$\mathcal{R}''_{B,1},\mathcal{R}'''_{B,1},\mathcal{R}''_{B,2},\mathcal{R}'''_{B,2}$
are defined in formulas (\ref{recurStep3B1}), (\ref{recurStep3B2}),
(\ref{recurStep3B3}) respectively, which are although products of three
lower-point tree amplitudes, but one of them is the three-point amplitude.
It is also important to notice how the forward singularities have been removed
in various terms by various methods.

%%%%%%%%%%%%%%%%%%%%%
\section{Some examples}
\label{secExample}
%%%%%%%%%%%%%%%%%%%%%%

In the previous section, we have presented a recursive formula for
one-loop integrand construction, based on the BCFW deformation and
$\mathcal{Q}$-cut construction. This new construction shows  that
there are other ways to
%the scale deformation as illustrated in the original
%$\mathcal{Q}$-cut construction \cite{Baadsgaard:2015twa} is not the
%only way of
write down a well-defined one-loop integrand which is valid up to
scale free terms.
%The BCFW deformation can also be introduced in the $\mathcal{Q}$-cut construction, although the result seems slightly complicated.
The recursive formula
(\ref{recurStepFinal}) has given an alternative factorization of
one-loop integrand, and it should be equivalent to the result of
original $\mathcal{Q}$-cut representation or Feynman diagram method,
at least up to some scale free terms. For a better understanding of
this recursive formula, in this section, we shall present detailed
computation of some one-loop integrands by recursive formula
(\ref{recurStepFinal}), and demonstrate their correspondence with
results of original $\mathcal{Q}$-cut construction and Feynman
diagram methods.

%%%%%%%%%%%%%%%
\subsection{The one-loop six-point amplitude in scalar $\phi^4$ theory}
%%%%%%%%%%%%%%%%%

In this example we consider the integrand of one-loop six-point
amplitude in color ordered scalar $\phi^4$ theory. For this theory, there is no
cubic vertex, so the computation is relatively simple since we do not
need to use the scale deformation to remove singular terms.  After using
appropriate BCFW deformation to get rid of boundary contribution, we
need to consider contributions from all detectable finite poles of
both $\mathcal{R}^{\mathcal{Q}}_{A}$ and
$\mathcal{R}^{\mathcal{Q}}_{B}$. In order to verify the equivalence
term by term, we will compute the integrand by Feynman diagram
method, the original $\cQ$-cut representation and the recursive
formula (\ref{recurStepFinal}).

\begin{figure}
  \centering
  \includegraphics[width=7in]{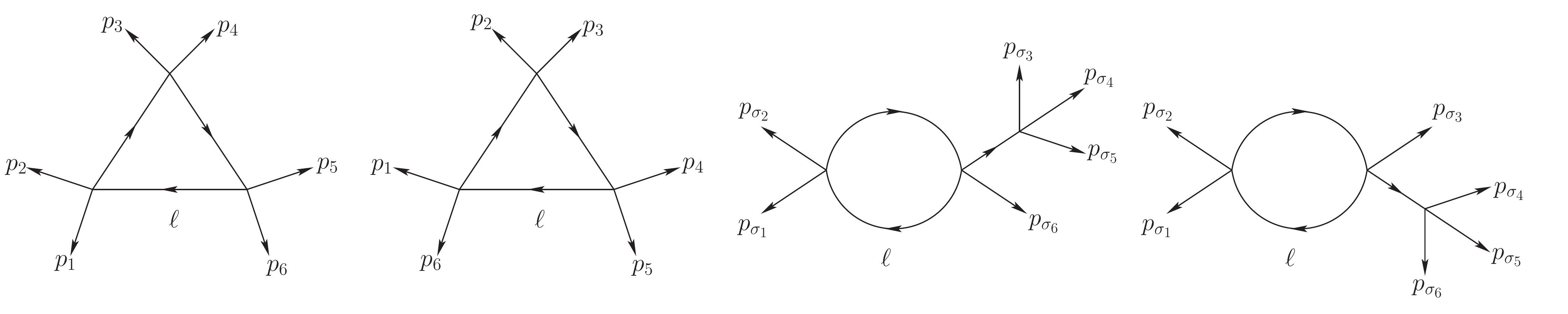}\\
  \caption{Feynman diagrams of color-ordered one-loop six-point amplitude in scalar $\phi^4$ theory.
  There are two triangle diagrams and twelve bubble diagrams with $\{\sigma_1,\ldots,\sigma_6\}\in \cyclic{1,2,3,4,5,6}$.}\label{FigsixPtPhi4FD}
\end{figure}

~\\~\\
{\bf Feynman diagram method}: there are in total fourteen Feynman
diagrams as shown in Figure \ref{FigsixPtPhi4FD}.
%%%%%%%%%%%%%%
%%%%%%%%%%%%%%%%
Using the Feynman rules, we directly get
\bea \cI^{\cF}&=&{1\over
\ell^2(\ell-p_{12})^2(\ell-p_{1234})^2}+{1\over
\ell^2(\ell-p_{61})^2(\ell-p_{6123})^2}\nonumber\\
&&+{1\over \ell^2(\ell-p_{\sigma_1\sigma_2})^2}{1\over
p_{\sigma_3\sigma_4\sigma_5}^2}+{1\over
\ell^2(\ell-p_{\sigma_1\sigma_2})^2}{1\over
p_{\sigma_4\sigma_5\sigma_6}^2}~~\mbox{for}~~\sigma\in\cyclic{1,2,3,4,5,6}~.~~~\eea
Applying the partial fraction identity
\bea {1\over D_1\cdots D_m}=\sum_{i=1}^{m}{1\over
D_i}\left[\prod_{j\neq i}{1\over D_j-D_i}\right]~,~~~\eea
we can rewrite above result as
\bea \cI^{\cF}&=&\Big\{{1\over \ell^2(-2\ell\cdot
p_{12}+p_{12}^2)(-2\ell\cdot p_{1234}+p_{1234}^2)}+\cyclic{1,2,3,4,5,6}\Big\}\nonumber\\
&&+\Big\{\Big({1\over \ell^2(-2\ell\cdot p_{12}+p_{12}^2)}+{1\over
\ell^2(-2\ell\cdot p_{3456}+p_{3456}^2)}\Big)\Big({1\over
p_{345}^2}+{1\over
p_{456}^2}\Big)+\cyclic{1,2,3,4,5,6}\Big\}~.~~~\label{FD6ptPhi4}\eea
When expanded, the first line contains 6 terms from triangle
diagrams, and the second line contains $4\times 6=24$ terms from
bubble diagrams.\\

~\\{\bf The $\cQ$-cut representation}: the integrand is given by
\bea \cI^{\cQ}&=&\cA_4(1,2,\WH{\ell}_R,-\WH{\ell}_L){1\over
\ell^2(-2\ell\cdot
p_{12}+p_{12}^2)}\cA_6(\WH{\ell}_L,-\WH{\ell}_R,3,4,5,6)\Big|_{\WH{\ell}=
\alpha_{12}\ell}+\cyclic{1,2,3,4,5,6}\nonumber\\
&&+\cA_6(1,2,3,4,\WH{\ell}_R,-\WH{\ell}_L){1\over \ell^2(-2\ell\cdot
p_{1234}+p_{1234}^2)}\cA_4(\WH{\ell}_L,-\WH{\ell}_R,5,6)\Big|_{\WH{\ell}=
\alpha_{1234}\ell}+\cyclic{1,2,3,4,5,6}~,~~~\nonumber\eea
where $\alpha_{i_1i_2}={p_{i_1i_2}^2\over 2\ell\cdot p_{i_1i_2}}$,
$\alpha_{i_1i_2i_3i_4}={p_{i_1i_2i_3i_4}^2\over 2\ell\cdot
p_{i_1i_2i_3i_4}}$ and $\ell^2=0$. The six-point tree-level
amplitude in general dimension is
\bea \cA_6(1,2,3,4,5,6)={1\over p_{123}^2}+{1\over
p_{234}^2}+{1\over p_{345}^2}~.~~~\eea
Inserting it back to above expression and rearranging some terms by
cyclic invariance, we get explicitly
\bea \cI^{\cQ}&=&\Big\{\Big({2\ell \cdot p_{12}\over
\ell^2(-2\ell\cdot p_{12}+p_{12}^2)}-{2\ell\cdot p_{1234}\over
\ell^2(-2\ell\cdot p_{1234}+p_{1234}^2)}\Big){1\over -(2\ell\cdot
p_{1234})p_{12}^2+(2\ell\cdot p_{12})p_{1234}^2}+\cyclic{1,2,3,4,5,6}\Big\}\nonumber\\
&&+\Big\{\Big({1\over \ell^2(-2\ell\cdot p_{12}+p_{12}^2)}+{1\over
\ell^2(-2\ell\cdot p_{1234}+p_{1234}^2)}\Big)\Big({1\over
p_{345}^2}+{1\over
p_{456}^2}\Big)+\cyclic{1,2,3,4,5,6}\Big\}~.~~~\label{Qcut6ptPhi4}\eea
The second line contains 24 terms, which is identical to the second
line of result (\ref{FD6ptPhi4}) by Feynman diagram method. The
first line contains 12 terms and can be organized as 6 pairs. The
sum of each pair leads to
\bea &&{(2\ell \cdot p_{12})(-2\ell\cdot
p_{1234}+p_{1234}^2)-(2\ell\cdot p_{1234})(-2\ell\cdot
p_{12}+p_{12}^2)\over \ell^2(-2\ell\cdot
p_{12}+p_{12}^2)(-2\ell\cdot p_{1234}+p_{1234}^2)}{1\over
-(2\ell\cdot p_{1234})p_{12}^2+(2\ell\cdot
p_{12})p_{1234}^2}+\cdots\nonumber\\
&=&{1\over \ell^2(-2\ell\cdot p_{12}+p_{12}^2)(-2\ell\cdot
p_{1234}+p_{1234}^2)}+\cyclic{1,2,3,4,5,6}~,~~~\eea
which equals to the 6 terms in the first line of result
(\ref{FD6ptPhi4}) by Feynman diagram method.\\

~\\{\bf Recursive formula}: now let us discuss the recursive
construction of $\cT^{\cQ}$ and the integrand $\cI={1\over
\ell^2}\cT^{\cQ}$. Because of the $\phi^4$ theory, in this example
only
$\mathcal{R}^{\mathcal{Q}}_{A,1},\mathcal{R}^{\mathcal{Q}}_{A,2}$
and $\cR'_{B,1},\cR'_{B,2}$ will contribute to the final integrand,
while the contributions
$\cR''_{B,1},\cR'''_{B,1},\cR''_{B,2},\cR'''_{B,2}$ are vanishing
since the three-point amplitude vanishes. Since we are considering
color-ordered amplitude, $\mathcal{T}^{\mathcal{Q}}$ will be the sum
of six diagrams,
\bea
\cT^{\cQ}&=&\cT^{\cQ}_1(\ell,-\ell,1,2,3,4,5,6)+\cT^{\cQ}_2(\ell,-\ell,2,3,4,5,6,1)+\cT^{\cQ}_3(\ell,-\ell,3,4,5,6,1,2)\nonumber\\
&&+\cT^{\cQ}_4(\ell,-\ell,4,5,6,1,2,3)+\cT^{\cQ}_5(\ell,-\ell,5,6,1,2,3,4)+\cT^{\cQ}_6(\ell,-\ell,6,1,2,3,4,5)~,~~~\eea
where in each diagram, one internal line has been cut. In order to
avoid boundary contribution, the two momenta to be deformed should
at least be separated by two legs. So we can take the BCFW
deformation as
\bea \WH{p}_1=p_1+zq~~,~~\WH{p}_4=p_4-zq~~,~~q^2=p_{1,4}\cdot
q=0~.~~~\eea
Note that we are not necessary to take the same deformation for all
$\mathcal{T}_i^\mathcal{Q}$'s. In the practical computation, we can
take the most convenient BCFW deformation for each
$\mathcal{T}_i^{\mathcal{Q}}$. But here we use the same deformation
for demonstration. Under this deformation, we then compute the
non-vanishing BCFW terms for each $\cT^{\cQ}_i$. Let us define
\bea &&z_{123}\equiv -{p_{123}^2\over 2q\cdot
p_{123}}~~,~~z_{561}\equiv -{p_{561}^2\over 2q\cdot
p_{561}}~~,~~z_{612}\equiv -{p_{612}^2\over 2q\cdot
p_{612}}~~,~~z_{12}^\pm\equiv -{\pm 2\ell\cdot p_{12}+p_{12}^2\over
2q\cdot(p_{12}\pm\ell)}~,~~~\\
&&z_{34}^\pm\equiv -{\pm 2\ell\cdot p_{34}+p_{34}^2\over
2q\cdot(p_{34}\pm\ell)}~~,~~z_{45}^\pm\equiv -{\pm 2\ell\cdot
p_{45}+p_{45}^2\over 2q\cdot(p_{45}\pm\ell)}~~,~~z_{61}^\pm\equiv
-{\pm 2\ell\cdot p_{61}+p_{61}^2\over
2q\cdot(p_{61}\pm\ell)}~.~~~\eea

For tree diagram of $\cT_1^{\cQ}$, there would be five contributing
terms under this deformation. The first is a
$\mathcal{R}^{\mathcal{Q}}_{A,2}$-type contribution,
\bea
\cT_{11}^{\cQ}&=&\cA_4(\WH{1},2,\WH{\ell}_R,-\WH{\ell}_L){1\over
-2\ell\cdot
p_{\WH{1}2}+p_{\WH{1}2}^2}\cA_4(\WH{\ell}_L,-\WH{\ell}_R,3,\WH{P}){1\over
p_{123}^2}\cA_4(-\WH{P},\WH{4},5,6)\nonumber\\
&=&{1\over -2\ell\cdot p_{\WH{1}2}+p_{\WH{1}2}^2}{1\over
p_{123}^2}\Big|_{z_{123}}~,~~~\eea
where $\WH{P}$ is understood to follow the momentum conservation of
each sub-amplitude, and $z=z_{123}$, $\alpha={p_{\WH{1}2}^2\over
2\ell\cdot p_{\WH{1}2}}\big|_{z_{123}}$. The second is a
$\mathcal{R}^{\mathcal{Q}}_{A,1}$-type contribution,
\bea \cT_{12}^{\cQ}&=&\cA_4(\WH{1},2,3,\WH{P}){1\over
p_{123}^2}\cA_4(-\WH{P},\WH{4},\WH{\ell}_R,-\WH{\ell}_L){1\over
-2\ell\cdot
p_{\WH{P}\WH{4}}+p_{\WH{P}\WH{4}}^2}\cA_4(\WH{\ell}_L,-\WH{\ell}_R,5,6)\nonumber\\
&=&{1\over p_{123}^2}{1\over 2\ell\cdot p_{56}+p_{56}^2}={1\over
-2\ell\cdot p_{1234}+p_{1234}^2}{1\over p_{123}^2}~,~~~\eea
where $z=z_{123}$, $\alpha=-{p_{56}^2\over 2\ell\cdot p_{56}}$. The
third is a $\mathcal{R}^{\mathcal{Q}}_{A,2}$-type contribution,
\bea
\cT_{13}^{\cQ}&=&\cA_4(\WH{1},2,\WH{\ell}_R,-\WH{\ell}_L){1\over
-2\ell\cdot
p_{\WH{1}2}+p_{\WH{1}2}^2}\cA_4(\WH{\ell}_L,-\WH{\ell}_R,\WH{P},6){1\over
p_{612}^2}\cA_4(-\WH{P},3,\WH{4},5)\nonumber\\
&=&{1\over -2\ell\cdot p_{\WH{1}2}+p_{\WH{1}2}^2}{1\over
p_{612}^2}\Big|_{z_{612}}~,~~~\eea
where $z=z_{612}$, $\alpha={p_{\WH{1}2}^2\over 2\ell\cdot
p_{\WH{1}2}}\big|_{z_{612}}$. The fourth is a
$\mathcal{R}^{\mathcal{Q}}_{A,1}$-type contribution,
\bea
\cT_{14}^{\cQ}&=&\cA_4(\WH{1},\WH{P},\WH{\ell}_R,-\WH{\ell}_L){1\over
-2\ell\cdot
p_{\WH{1}\WH{P}}+p_{\WH{1}\WH{P}}^2}\cA_4(\WH{\ell}_L,-\WH{\ell}_R,5,6){1\over
p_{561}^2}\cA_4(-\WH{P},2,3,\WH{4})\nonumber\\
&=&{1\over 2\ell\cdot p_{56}+p_{56}^2}{1\over p_{561}^2}={1\over
-2\ell\cdot p_{1234}+p_{1234}^2}{1\over p_{561}^2}~,~~~\eea
where $z=z_{561}$, $\alpha=-{p_{56}^2\over 2\ell\cdot p_{56}}$.
Finally, the fifth is a $\mathcal{R}'_{B,2}$ contribution,
\bea \cT_{15}^{\cQ}&=&\cA_4(-\ell,\WH{1},2,\WH{P}){1\over
(p_{12}-\ell)^2}\cA_6(-\WH{P},3,\WH{4},5,6,\ell)={1\over
(p_{12}-\ell)^2}\Big({1\over p_{3\WH{4}5}^2}+{1\over
p_{\WH{4}56}^2}+{1\over
(\ell+p_{56})^2}\Big)\Big|_{\ell^2=0,z=z_{12}^-}\nonumber\\
&=&{1\over -2\ell\cdot p_{12}+p_{12}^2 }{1\over
p_{3\WH{4}5}^2}\Big|_{z_{12}^-}+{1\over -2\ell\cdot p_{12}+p_{12}^2
}{1\over p_{\WH{4}56}^2}\Big|_{z_{12}^-}+{1\over -2\ell\cdot
p_{12}+p_{12}^2 }{1\over -2\ell\cdot
p_{1234}+p_{1234}^2}\nonumber\\
&\equiv&\cT_{15,1}^{\cQ}+\cT_{15,2}^{\cQ}+\cT_{15,3}^{\cQ} ~,~~~\eea
where $z=z_{12}^-$.

So for $\mathcal{T}_{1}^{\mathcal{Q}}$, in total we get seven terms.
Let us see how these seven terms is corresponding to the terms in
$\mathcal{Q}$-cut representation. $\mathcal{T}^{\mathcal{Q}}_{12}$,
$\mathcal{T}^{\mathcal{Q}}_{14}$ and
$\mathcal{T}^{\mathcal{Q}}_{15,3}$ are evaluated with the
un-deformed momenta. It is simple to see that ${1\over
\ell^2}\mathcal{T}^{\mathcal{Q}}_{15,3}$ corresponds to a term in
the first line of (\ref{FD6ptPhi4}), while ${1\over
\ell^2}\mathcal{T}^{\mathcal{Q}}_{12}$, ${1\over
\ell^2}\mathcal{T}^{\mathcal{Q}}_{14}$ also have their equivalent
terms in the second line of (\ref{FD6ptPhi4}),
\bea {1\over
\ell^2}(\mathcal{T}^{\mathcal{Q}}_{12}+\mathcal{T}^{\mathcal{Q}}_{14})={1\over\ell^2(
-2\ell\cdot p_{1234}+p_{1234}^2)}\left({1\over p_{123}^2}+{1\over
p_{234}^2}\right)~.~~~\eea
There are also four terms $\mathcal{T}^{\mathcal{Q}}_{11}$,
$\mathcal{T}^{\mathcal{Q}}_{13}$,
$\mathcal{T}^{\mathcal{Q}}_{15,1}$,
$\mathcal{T}^{\mathcal{Q}}_{15,2}$ evaluated with deformed momenta.
We have
\bea &&\cT^{\cQ}_{11}+\cT^{\cQ}_{15,2}={1\over -2\ell\cdot
p_{\WH{1}2}+p_{\WH{1}2}^2}{1\over p_{456}^2}\Big|_{z_{123}}+{1\over
-2\ell\cdot p_{12}+p_{12}^2
}{1\over p_{\WH{4}56}^2}\Big|_{z_{12}^-}\nonumber\\
&&={1\over p_{456}^2}{1\over (-2\ell\cdot p_{12}+p_{12}^2)+{(2q\cdot
p_{12}-2q\cdot \ell)\over 2q\cdot p_{456}}p_{456}^2}+{1\over
(-2\ell\cdot p_{12}+p_{12}^2)}{1\over p_{456}^2+{2q\cdot
p_{456}\over(2q\cdot p_{12}-2q\cdot \ell) }(-2\ell\cdot
p_{12}+p_{12}^2)}~,~~~\nonumber\eea
as well as
\bea &&\cT^{\cQ}_{13}+\cT^{\cQ}_{15,1}={1\over -2\ell\cdot
p_{\WH{1}2}+p_{\WH{1}2}^2}{1\over p_{345}^2}\Big|_{z_{612}}+{1\over
-2\ell\cdot p_{12}+p_{12}^2
}{1\over p_{3\WH{4}5}^2}\Big|_{z_{12}^-}\nonumber\\
&&={1\over p_{345}^2}{1\over (-2\ell\cdot p_{12}+p_{12}^2)+{(2q\cdot
p_{12}-2q\cdot \ell)\over 2q\cdot p_{345}}p_{345}^2}+{1\over
(-2\ell\cdot p_{12}+p_{12}^2)}{1\over p_{345}^2+{2q\cdot
p_{345}\over(2q\cdot p_{12}-2q\cdot \ell) }(-2\ell\cdot
p_{12}+p_{12}^2)}~.~~~\nonumber\eea
Using the identity
\bea {1\over A(B-\lambda A)}+{1\over B(A-{1\over \lambda}
B)}={1\over AB}~,~~~\label{idenAB}\eea
we arrive at
\bea {1\over
\ell^2}(\cT^{\cQ}_{11}+\cT^{\cQ}_{15,2}+\cT^{\cQ}_{13}+\cT^{\cQ}_{15,1})={1\over
\ell^2(-2\ell\cdot p_{12}+p_{12}^2)}\Big({1\over p_{345}^2}+{1\over
p_{456}^2}\Big)~.~~~\eea
The above computation shows the one-to-one correspondence between
the results of Feynman diagram method and the recursive formula. The
contribution of ${1\over \ell^2}\mathcal{T}_{1}^{\mathcal{Q}}$ is
equivalent to the terms in (\ref{FD6ptPhi4}) with a specific cyclic
permutation.

Similarly, we can also check the equivalence of the other five
$\mathcal{T}_{i}^{\mathcal{Q}}$ with the terms in (\ref{FD6ptPhi4})
of the other cyclic permutation. For tree diagram of $\cT_2^{\cQ}$,
there would also be five contributing terms. The first is a
$\mathcal{R}^{\mathcal{Q}}_{A,2}$-type contribution,
{\small \bea
\cT_{21}^{\cQ}&=&\cA_4(2,3,\WH{\ell}_R,-\WH{\ell}_L){1\over
-2\ell\cdot
p_{23}+p_{23}^2}\cA_4(\WH{\ell}_L,-\WH{\ell}_R,\WH{P},\WH{1}){1\over
p_{123}^2}\cA_4(-\WH{P},\WH{4},5,6) ={1\over -2\ell\cdot
p_{23}+p_{23}^2}{1\over p_{123}^2}~,~~~\nonumber\eea}
where $z=z_{123}$, $\alpha={p_{23}^2\over 2\ell\cdot p_{23}}$. The
second is a $\mathcal{R}^{\mathcal{Q}}_{A,2}$-type contribution,
{\small \bea
\cT_{22}^{\cQ}&=&\cA_4(2,\WH{P},\WH{\ell}_R,-\WH{\ell}_L){1\over
-2\ell\cdot
p_{2\WH{P}}+p_{2\WH{P}}^2}\cA_4(\WH{\ell}_L,-\WH{\ell}_R,6,\WH{1}){1\over
p_{612}^2}\cA_4(-\WH{P},3,\WH{4},5)={1\over -2\ell\cdot
p_{23\WH{4}5}+p_{23\WH{4}5}^2}{1\over
p_{612}^2}\Big|_{z_{612}}~,~~~\nonumber\eea}
where $z=z_{612}$, $\alpha=-{p_{6\WH{1}}^2\over 2\ell\cdot
p_{6\WH{1}}}\big|_{z_{612}}$. The third is a
$\mathcal{R}^{\mathcal{Q}}_{A,2}$-type contribution,
{\small \bea
\cT_{23}^{\cQ}&=&\cA_4(\WH{P},5,\WH{\ell}_R,-\WH{\ell}_L){1\over
-2\ell\cdot
p_{5\WH{P}}+p_{5\WH{P}}^2}\cA_4(\WH{\ell}_L,-\WH{\ell}_R,6,\WH{1}){1\over
p_{561}^2}\cA_4(-\WH{P},2,3,\WH{4})={1\over -2\ell\cdot
p_{23\WH{4}5}+p_{23\WH{4}5}^2}{1\over
p_{561}^2}\Big|_{z_{561}}~,~~~\nonumber\eea}
where $z=z_{561}$, $\alpha=-{p_{6\WH{1}}^2\over 2\ell\cdot
p_{6\WH{1}}}\big|_{z_{561}}$. The fourth is a
$\mathcal{R}^{\mathcal{Q}}_{A,1}$-type contribution,
{\small\bea \cT_{24}^{\cQ}&=&\cA_4(5,6,\WH{1},\WH{P}){1\over
p_{561}^2}\cA_4(2,3,\WH{\ell}_R,-\WH{\ell}_L){1\over -2\ell\cdot
p_{23}+p_{23}^2}\cA_4(\WH{\ell}_L,-\WH{\ell}_R,\WH{4},-\WH{P})={1\over
p_{561}^2}{1\over -2\ell\cdot p_{23}+p_{23}^2}~,~~~\nonumber\eea}
where $z=z_{561}$, $\alpha={p_{23}^2\over 2\ell\cdot p_{23}}$.
Finally the fifth is a $\mathcal{R}'_{B,1}$-type contribution,
{\small \bea \cT_{25}^{\cQ}&=&\cA_4(6,\WH{1},\ell,\WH{P}){1\over
(p_{61}+\ell)^2}\cA_6(-\WH{P},-\ell,2,3,\WH{4},5)={1\over
(p_{61}+\ell)^2}\Big({1\over p_{23\WH{4}}^2}+{1\over
p_{3\WH{4}5}^2}+{1\over
(-\ell+p_{23})^2}\Big)\Big|_{\ell^2=0,z=z_{61}^+}\nonumber\\
&=&{1\over -2\ell\cdot p_{2345}+p_{2345}^2 }{1\over
p_{23\WH{4}}^2}\Big|_{z_{61}^+}+{1\over -2\ell\cdot
p_{2345}+p_{2345}^2 }{1\over p_{3\WH{4}5}^2}\Big|_{z_{61}^+}+{1\over
-2\ell\cdot p_{2345}+p_{2345}^2 }{1\over -2\ell\cdot
p_{23}+p_{23}^2}\nonumber\\
&\equiv&\cT_{25,1}^{\cQ}+\cT_{25,2}^{\cQ}+\cT_{25,3}^{\cQ}
~,~~~\nonumber\eea}
where $z=z_{61}^+$.

For tree diagrams of $\cT_3^{\cQ}$, there are in total six
contributing terms. The first is a
$\mathcal{R}^{\mathcal{Q}}_{A,2}$-type contribution,
{\small \bea
\cT_{31}^{\cQ}&=&\cA_4(3,\WH{P},\WH{\ell}_R,-\WH{\ell}_L){1\over
-2\ell\cdot
p_{3\WH{P}}+p_{3\WH{P}}^2}\cA_4(\WH{\ell}_L,-\WH{\ell}_R,\WH{1},2){1\over
p_{123}^2}\cA_4(-\WH{P},\WH{4},5,6)={1\over -2\ell\cdot
p_{3\WH{4}56}+p_{3\WH{4}56}^2}{1\over
p_{123}^2}\Big|_{z_{123}}~,~~~\nonumber\eea}
where $z=z_{123}$, $\alpha=-{p_{\WH{1}2}^2\over 2\ell\cdot
p_{\WH{1}2}}\big|_{z_{123}}$. The second is a
$\mathcal{R}^{\mathcal{Q}}_{A,2}$-type contribution,
{\small \bea
\cT_{32}^{\cQ}&=&\cA_4(\WH{P},6,\WH{\ell}_R,-\WH{\ell}_L){1\over
-2\ell\cdot
p_{\WH{P}6}+p_{\WH{P}6}^2}\cA_4(\WH{\ell}_L,-\WH{\ell}_R,\WH{1},2){1\over
p_{612}^2}\cA_4(-\WH{P},3,\WH{4},5)={1\over -2\ell\cdot
p_{3\WH{4}56}+p_{3\WH{4}56}^2}{1\over
p_{612}^2}\Big|_{z_{612}}~,~~~\nonumber\eea}
where $z=z_{612}$, $\alpha=-{p_{\WH{1}2}^2\over 2\ell\cdot
p_{\WH{1}2}}\big|_{z_{612}}$. The third is a
$\mathcal{R}^{\mathcal{Q}}_{A,1}$-type contribution,
{\small \bea \cT_{33}^{\cQ}&=&\cA_4(6,\WH{1},2,\WH{P}){1\over
p_{612}^2}\cA_4(3,\WH{4},\WH{\ell}_R,-\WH{\ell}_L){1\over
-2\ell\cdot
p_{3\WH{4}}+p_{3\WH{4}}^2}\cA_4(\WH{\ell}_L,-\WH{\ell}_R,5,-\WH{P})={1\over
p_{612}^2}{1\over -2\ell\cdot
p_{3\WH{4}}+p_{3\WH{4}}^2}\Big|_{z_{612}}~,~~~\nonumber\eea}
where $z=z_{612}$, $\alpha={p_{3\WH{4}}^2\over 2\ell\cdot
p_{3\WH{4}}}\big|_{z_{612}}$. The fourth is a
$\mathcal{R}^{\mathcal{Q}}_{A,1}$-type contribution,
{\small \bea \cT_{34}^{\cQ}&=&\cA_4(5,6,\WH{1},\WH{P}){1\over
p_{561}^2}\cA_4(3,\WH{4},\WH{\ell}_R,-\WH{\ell}_L){1\over
-2\ell\cdot
p_{3\WH{4}}+p_{3\WH{4}}^2}\cA_4(\WH{\ell}_L,-\WH{\ell}_R,-\WH{P},2)={1\over
p_{561}^2}{1\over -2\ell\cdot
p_{3\WH{4}}+p_{3\WH{4}}^2}\Big|_{z_{561}}~,~~~\nonumber\eea}
where $z=z_{561}$, $\alpha={p_{3\WH{4}}^2\over 2\ell\cdot
p_{3\WH{4}}}\big|_{z_{561}}$. The fifth is a
$\mathcal{R}'_{B,1}$-type contribution,
{\small \bea \cT_{35}^{\cQ}&=&\cA_4(\WH{1},2,\ell,\WH{P}){1\over
(p_{12}+\ell)^2}\cA_6(-\WH{P},-\ell,3,\WH{4},5,6)={1\over
(p_{12}+\ell)^2}\Big({1\over p_{3\WH{4}5}^2}+{1\over
p_{\WH{4}56}^2}+{1\over
(-\ell+p_{3\WH{4}})^2}\Big)\Big|_{\ell^2=0,z=z_{12}^+}\nonumber\\
&=&{1\over -2\ell\cdot p_{3456}+p_{3456}^2 }{1\over
p_{3\WH{4}5}^2}\Big|_{z_{12}^+}+{1\over -2\ell\cdot
p_{3456}+p_{3456}^2 }{1\over p_{\WH{4}56}^2}\Big|_{z_{12}^+}+{1\over
-2\ell\cdot p_{3456}+p_{3456}^2 }{1\over -2\ell\cdot
p_{3\WH{4}}+p_{3\WH{4}}^2}\Big|_{z_{12}^+}\nonumber\\
&\equiv&\cT_{35,1}^{\cQ}+\cT_{35,2}^{\cQ}+\cT_{35,3}^{\cQ}
~,~~~\nonumber\eea}
where $z=z_{12}^+$. Finally the sixth is a $\mathcal{R}'_{B,1}$-type
contribution,
{\small \bea \cT_{36}^{\cQ}&=&\cA_6(5,6,\WH{1},2,\ell,\WH{P}){1\over
(p_{5612}+\ell)^2}\cA_4(-\WH{P},-\ell,3,\WH{4})=\Big({1\over
p_{56\WH{1}}^2}+{1\over p_{6\WH{1}2}^2}+{1\over
(\ell+p_{\WH{1}2})^2}\Big){1\over
(p_{5612}+\ell)^2}\Big|_{\ell^2=0,z=-z_{34}^-}\nonumber\\
&=&{1\over -2\ell\cdot p_{34}+p_{34}^2 }{1\over
p_{56\WH{1}}^2}\Big|_{-z_{34}^-}+{1\over -2\ell\cdot p_{34}+p_{34}^2
}{1\over p_{6\WH{1}2}^2}\Big|_{-z_{34}^-}+{1\over -2\ell\cdot
p_{34}+p_{34}^2 }{1\over -2\ell\cdot
p_{3\WH{4}56}+p_{3\WH{4}56}^2}\Big|_{-z_{34}^-}\nonumber\\
&\equiv&\cT_{36,1}^{\cQ}+\cT_{36,2}^{\cQ}+\cT_{36,3}^{\cQ}
~,~~~\eea}
where $z=-z_{34}^-$.

For tree diagrams of $\cT_4^{\cQ}$, there are in total five
contributing terms. The first is a
$\mathcal{R}^{\mathcal{Q}}_{A,2}$-type contribution,
{\small \bea
\cT_{41}^{\cQ}&=&\cA_4(\WH{P},\WH{1},\WH{\ell}_R,-\WH{\ell}_L){1\over
-2\ell\cdot
p_{\WH{1}\WH{P}}+p_{\WH{1}\WH{P}}^2}\cA_4(\WH{\ell}_L,-\WH{\ell}_R,2,3){1\over
p_{123}^2}\cA_4(-\WH{P},\WH{4},5,6)={1\over -2\ell\cdot
p_{4561}+p_{4561}^2}{1\over p_{123}^2}~,~~~\nonumber\eea}
where $z=z_{123}$, $\alpha=-{p_{23}^2\over 2\ell\cdot p_{23}}$. The
second is a $\mathcal{R}^{\mathcal{Q}}_{A,1}$-type contribution,
{\small \bea \cT_{42}^{\cQ}&=&\cA_4(\WH{1},2,3,\WH{P}){1\over
p_{123}^2}\cA_4(\WH{4},5,\WH{\ell}_R,-\WH{\ell}_L){1\over
-2\ell\cdot
p_{\WH{4}5}+p_{\WH{4}5}^2}\cA_4(\WH{\ell}_L,-\WH{\ell}_R,6,-\WH{P})={1\over
p_{123}^2}{1\over -2\ell\cdot
p_{\WH{4}5}+p_{\WH{4}5}^2}\Big|_{z_{123}}~,~~~\nonumber\eea}
where $z=z_{123}$, $\alpha={p_{\WH{4}5}^2\over 2\ell\cdot
p_{\WH{4}5}}\big|_{z_{123}}$. The third is a
$\mathcal{R}^{\mathcal{Q}}_{A,1}$-type contribution,
{\small \bea \cT_{43}^{\cQ}&=&\cA_4(6,\WH{1},2,\WH{P}){1\over
p_{612}^2}\cA_4(\WH{4},5,\WH{\ell}_R,-\WH{\ell}_L){1\over
-2\ell\cdot
p_{\WH{4}5}+p_{\WH{4}5}^2}\cA_4(\WH{\ell}_L,-\WH{\ell}_R,-\WH{P},3)={1\over
p_{612}^2}{1\over -2\ell\cdot
p_{\WH{4}5}+p_{\WH{4}5}^2}\Big|_{z_{612}}~,~~~\nonumber\eea}
where $z=z_{612}$, $\alpha={p_{\WH{4}5}^2\over 2\ell\cdot
p_{\WH{4}5}}\big|_{z_{612}}$. The fourth is a
$\mathcal{R}^{\mathcal{Q}}_{A,1}$-type contribution,
{\small \bea \cT_{44}^{\cQ}&=&\cA_4(5,6,\WH{1},\WH{P}){1\over
p_{561}^2}\cA_4(\WH{4},-\WH{P},\WH{\ell}_R,-\WH{\ell}_L){1\over
-2\ell\cdot
p_{\WH{4}\WH{P}}+p_{\WH{4}\WH{P}}^2}\cA_4(\WH{\ell}_L,-\WH{\ell}_R,2,3)={1\over
p_{561}^2}{1\over -2\ell\cdot
p_{4561}+p_{4561}^2}~,~~~\nonumber\eea}
where $z=z_{561}$, $\alpha=-{p_{23}^2\over 2\ell\cdot p_{23}}$. The
fifth is a $\mathcal{R}'_{B,1}$-type contribution,
{\small \bea \cT_{45}^{\cQ}&=&\cA_6(6,\WH{1},2,3,\ell,\WH{P}){1\over
(p_{6123}+\ell)^2}\cA_4(-\WH{P},-\ell,\WH{4},5)=\Big({1\over
p_{6\WH{1}2}^2}+{1\over p_{\WH{1}23}^2}+{1\over
(\ell+p_{23})^2}\Big){1\over
(p_{6123}+\ell)^2}\Big|_{\ell^2=0,z=-z_{45}^-}\nonumber\\
&=&{1\over -2\ell\cdot p_{45}+p_{45}^2 }{1\over
p_{6\WH{1}2}^2}\Big|_{-z_{45}^-}+{1\over -2\ell\cdot p_{45}+p_{45}^2
}{1\over p_{\WH{1}23}^2}\Big|_{-z_{45}^-}+{1\over -2\ell\cdot
p_{45}+p_{45}^2 }{1\over -2\ell\cdot
p_{4561}+p_{4561}^2}\nonumber\\
&\equiv&\cT_{45,1}^{\cQ}+\cT_{45,2}^{\cQ}+\cT_{45,3}^{\cQ}
~,~~~\nonumber\eea}
where $z=-z_{45}^-$.

For tree diagram of $\cT_5^{\cQ}$, there are in total five
contributing terms. The first is a
$\mathcal{R}^{\mathcal{Q}}_{A,1}$-type contribution,
{\small \bea \cT_{51}^{\cQ}&=&\cA_4(\WH{1},2,3,\WH{P}){1\over
p_{123}^2}\cA_4(5,6,\WH{\ell}_R,-\WH{\ell}_L){1\over -2\ell\cdot
p_{56}+p_{56}^2}\cA_4(\WH{\ell}_L,-\WH{\ell}_R,-\WH{P},\WH{4})={1\over
p_{123}^2}{1\over -2\ell\cdot p_{56}+p_{56}^2}~,~~~\nonumber\eea}
where $z=z_{123}$, $\alpha={p_{56}^2\over 2\ell\cdot p_{56}}$. The
second is a $\mathcal{R}^{\mathcal{Q}}_{A,1}$-type contribution,
{\small \bea \cT_{52}^{\cQ}&=&\cA_4(6,\WH{1},2,\WH{P}){1\over
p_{612}^2}\cA_4(5,-\WH{P},\WH{\ell}_R,-\WH{\ell}_L){1\over
-2\ell\cdot
p_{5\WH{P}}+p_{5\WH{P}}^2}\cA_4(\WH{\ell}_L,-\WH{\ell}_R,3,\WH{4})={1\over
p_{612}^2}{1\over -2\ell\cdot
p_{56\WH{1}2}+p_{56\WH{1}2}^2}\Big|_{z_{612}}~,~~~\nonumber\eea}
where $z=z_{612}$, $\alpha=-{p_{3\WH{4}}^2\over 2\ell\cdot
p_{3\WH{4}}}\big|_{z_{612}}$. The third is a
$\mathcal{R}^{\mathcal{Q}}_{A,1}$-type contribution,
{\small \bea \cT_{53}^{\cQ}&=&\cA_4(5,6,\WH{1},\WH{P}){1\over
p_{561}^2}\cA_4(-\WH{P},2,\WH{\ell}_R,-\WH{\ell}_L){1\over
-2\ell\cdot
p_{2\WH{P}}+p_{2\WH{P}}^2}\cA_4(\WH{\ell}_L,-\WH{\ell}_R,3,\WH{4})={1\over
p_{561}^2}{1\over -2\ell\cdot
p_{56\WH{1}2}+p_{56\WH{1}2}^2}\Big|_{z_{561}}~,~~~\nonumber\eea}
where $z=z_{561}$, $\alpha=-{p_{3\WH{4}}^2\over 2\ell\cdot
p_{3\WH{4}}}\big|_{z_{561}}$. The fourth is a
$\mathcal{R}^{\mathcal{Q}}_{A,2}$-type contribution,
{\small \bea
\cT_{54}^{\cQ}&=&\cA_4(5,6,\WH{\ell}_R,-\WH{\ell}_L){1\over
-2\ell\cdot
p_{56}+p_{56}^2}\cA_4(\WH{\ell}_L,-\WH{\ell}_R,\WH{1},\WH{P}){1\over
p_{561}^2}\cA_4(-\WH{P},2,3,\WH{4})={1\over -2\ell\cdot
p_{56}+p_{56}^2}{1\over p_{561}^2}~,~~~\nonumber\eea}
where $z=z_{561}$, $\alpha={p_{56}^2\over 2\ell\cdot p_{56}}$. The
fifth is a $\mathcal{R}'_{B,2}$-type contribution,
{\small \bea
\cT_{55}^{\cQ}&=&\cA_6(-\ell,5,6,\WH{1},2,\WH{P}){1\over
(p_{5612}-\ell)^2}\cA_4(-\WH{P},3,\WH{4},\ell)=\Big({1\over
p_{56\WH{1}}^2}+{1\over p_{6\WH{1}2}^2}+{1\over
(p_{56}-\ell)^2}\Big){1\over
(p_{5612}-\ell)^2}\Big|_{\ell^2=0,z=-z_{34}^+}\nonumber\\
&=&{1\over -2\ell\cdot p_{5612}+p_{5612}^2 }{1\over
p_{56\WH{1}}^2}\Big|_{-z_{34}^+}+{1\over -2\ell\cdot
p_{5612}+p_{5612}^2 }{1\over
p_{6\WH{1}2}^2}\Big|_{-z_{34}^+}+{1\over -2\ell\cdot
p_{5612}+p_{5612}^2 }{1\over -2\ell\cdot
p_{56}+p_{56}^2}\nonumber\\
&\equiv&\cT_{55,1}^{\cQ}+\cT_{55,2}^{\cQ}+\cT_{55,3}^{\cQ}
~,~~~\nonumber\eea}
where $z=-z_{34}^+$.

For tree diagrams of $\cT_6^{\cQ}$, there are in total six
contributing terms. The first is a
$\mathcal{R}^{\mathcal{Q}}_{A,1}$-type contribution,
{\small \bea \cT_{61}^{\cQ}&=&\cA_4(\WH{1},2,3,\WH{P}){1\over
p_{123}^2}\cA_4(6,-\WH{P},\WH{\ell}_R,-\WH{\ell}_L){1\over
-2\ell\cdot
p_{6\WH{P}}+p_{6\WH{P}}^2}\cA_4(\WH{\ell}_L,-\WH{\ell}_R,\WH{4},5)={1\over
p_{123}^2}{1\over -2\ell\cdot
p_{6\WH{1}23}+p_{6\WH{1}23}^2}\Big|_{z_{123}}~,~~~\nonumber\eea}
where $z=z_{123}$, $\alpha=-{p_{\WH{4}5}^2\over 2\ell\cdot
p_{\WH{4}5}}\big|_{z_{123}}$. The second is a
$\mathcal{R}^{\mathcal{Q}}_{A,1}$-type contribution,
{\small \bea \cT_{62}^{\cQ}&=&\cA_4(6,\WH{1},2,\WH{P}){1\over
p_{612}^2}\cA_4(-\WH{P},3,\WH{\ell}_R,-\WH{\ell}_L){1\over
-2\ell\cdot
p_{3\WH{P}}+p_{3\WH{P}}^2}\cA_4(\WH{\ell}_L,-\WH{\ell}_R,\WH{4},5)={1\over
p_{612}^2}{1\over -2\ell\cdot
p_{6\WH{1}23}+p_{6\WH{1}23}^2}\Big|_{z_{612}}~,~~~\nonumber\eea}
where $z=z_{612}$, $\alpha=-{p_{\WH{4}5}^2\over 2\ell\cdot
p_{\WH{4}5}}\big|_{z_{612}}$. The third is a
$\mathcal{R}^{\mathcal{Q}}_{A,2}$-type contribution,
{\small \bea
\cT_{63}^{\cQ}&=&\cA_4(6,\WH{1},\WH{\ell}_R,-\WH{\ell}_L){1\over
-2\ell\cdot
p_{6\WH{1}}+p_{6\WH{1}}^2}\cA_4(\WH{\ell}_L,-\WH{\ell}_R,2,\WH{P}){1\over
p_{612}^2}\cA_4(-\WH{P},3,\WH{4},5)={1\over -2\ell\cdot
p_{6\WH{1}}+p_{6\WH{1}}^2}{1\over
p_{612}^2}\Big|_{z_{612}}~,~~~\nonumber\eea}
where $z=z_{612}$, $\alpha={p_{6\WH{1}}^2\over 2\ell\cdot
p_{6\WH{1}}}\big|_{z_{612}}$. The fourth is a
$\mathcal{R}^{\mathcal{Q}}_{A,2}$-type contribution,
{\small \bea
\cT_{64}^{\cQ}&=&\cA_4(6,\WH{1},\WH{\ell}_R,-\WH{\ell}_L){1\over
-2\ell\cdot
p_{6\WH{1}}+p_{6\WH{1}}^2}\cA_4(\WH{\ell}_L,-\WH{\ell}_R,\WH{P},5){1\over
p_{561}^2}\cA_4(-\WH{P},2,3,\WH{4})={1\over -2\ell\cdot
p_{6\WH{1}}+p_{6\WH{1}}^2}{1\over
p_{561}^2}\Big|_{z_{561}}~,~~~\nonumber\eea}
where $z=z_{561}$, $\alpha={p_{6\WH{1}}^2\over 2\ell\cdot
p_{6\WH{1}}}\big|_{z_{561}}$. The fifth is a
$\mathcal{R}'_{B,2}$-type contribution,
{\small \bea \cT_{65}^{\cQ}&=&\cA_4(-\ell,6,\WH{1},\WH{P}){1\over
(p_{61}-\ell)^2}\cA_6(-\WH{P},2,3,\WH{4},5,\ell)={1\over
(p_{61}-\ell)^2}\Big({1\over p_{23\WH{4}}^2}+{1\over
p_{3\WH{4}5}^2}+{1\over
(\ell+p_{\WH{4}5})^2}\Big)\Big|_{\ell^2=0,z=z_{61}^-}\nonumber\\
&=&{1\over -2\ell\cdot p_{61}+p_{61}^2 }{1\over
p_{23\WH{4}}^2}\Big|_{z_{61}^-}+{1\over -2\ell\cdot p_{61}+p_{61}^2
}{1\over p_{3\WH{4}5}^2}\Big|_{z_{61}^-}+{1\over -2\ell\cdot
p_{61}+p_{61}^2 }{1\over -2\ell\cdot
p_{6\WH{1}23}+p_{6\WH{1}23}^2}\Big|_{z_{61}^-}\nonumber\\
&\equiv&\cT_{65,1}^{\cQ}+\cT_{65,2}^{\cQ}+\cT_{65,3}^{\cQ}
~,~~~\nonumber\eea}
where $z=z_{61}^-$. Finally, the sixth is a
$\mathcal{R}'_{B,2}$-type contribution,
{\small \bea
\cT_{66}^{\cQ}&=&\cA_6(-\ell,6,\WH{1},2,3,\WH{P}){1\over
(p_{6123}-\ell)^2}\cA_4(-\WH{P},\WH{4},5,\ell)=\Big({1\over
p_{6\WH{1}2}^2}+{1\over p_{\WH{1}23}^2}+{1\over
(-\ell+p_{6\WH{1}})^2}\Big){1\over
(p_{6123}-\ell)^2}\Big|_{\ell^2=0,z=-z_{45}^+}\nonumber\\
&=&{1\over -2\ell\cdot p_{6123}+p_{6123}^2 }{1\over
p_{6\WH{1}2}^2}\Big|_{-z_{45}^+}+{1\over -2\ell\cdot
p_{6123}+p_{6123}^2 }{1\over
p_{\WH{1}23}^2}\Big|_{-z_{45}^+}+{1\over -2\ell\cdot
p_{6123}+p_{6123}^2 }{1\over -2\ell\cdot
p_{6\WH{1}}+p_{6\WH{1}}^2}\Big|_{-z_{45}^+}\nonumber\\
&\equiv&\cT_{66,1}^{\cQ}+\cT_{66,2}^{\cQ}+\cT_{66,3}^{\cQ}
~,~~~\nonumber\eea}
where $z=-z_{45}^+$.

All the above results in total generate 48 terms. As is done for
$\mathcal{T}^{\mathcal{Q}}_{1}$, it can be checked that, the 4 terms
with un-deformed momenta
\bea &&{1\over
\ell^2}(\cT^{\cQ}_{15,3}+\cT^{\cQ}_{25,3}+\cT^{\cQ}_{45,3}+\cT^{\cQ}_{55,3})\\
&=&{1\over \ell^2(-2\ell\cdot p_{12}+p_{12}^2)(-2\ell\cdot
p_{1234}+p_{1234}^2)}+{1\over \ell^2(-2\ell\cdot
p_{23}+p_{23}^2)(-2\ell\cdot p_{2345}+p_{2345}^2)}\nonumber\\
&&+{1\over \ell^2(-2\ell\cdot p_{45}+p_{45}^2)(-2\ell\cdot
p_{4561}+p_{4561}^2)}+{1\over \ell^2(-2\ell\cdot
p_{56}+p_{56}^2)(-2\ell\cdot p_{5612}+p_{5612}^2)}~~~~\nonumber\eea
reproduce the 4 four terms in the first line of (\ref{FD6ptPhi4}).
While
\bea {1\over \ell^2}(\cT^{\cQ}_{35,3}+\cT^{\cQ}_{36,3})={1\over
\ell^2(-2\ell\cdot p_{34}+p_{34}^2)(-2\ell\cdot
p_{3456}+p_{3456}^2)}~~~~\eea
and
\bea {1\over \ell^2}(\cT^{\cQ}_{65,3}+\cT^{\cQ}_{66,3})={1\over
\ell^2(-2\ell\cdot p_{61}+p_{61}^2)(-2\ell\cdot
p_{6123}+p_{6123}^2)}~~~~\eea
reproduce the other 2 in the first line of (\ref{FD6ptPhi4}).

For the comparison of the second line in (\ref{FD6ptPhi4}), we have
\bea &&{1\over
\ell^2}(\cT^{\cQ}_{21}+\cT^{\cQ}_{24}+\cT^{\cQ}_{41}+\cT^{\cQ}_{44}+\cT^{\cQ}_{51}+\cT^{\cQ}_{54}+\cT^{\cQ}_{12}+\cT^{\cQ}_{14})\\
&&~~~~~~=\Big({1\over \ell^2(-2\ell\cdot p_{23}+p_{23}^2)}+{1\over
\ell^2(-2\ell\cdot p_{4561}+p_{4561}^2)}\Big)\Big({1\over
p_{456}^2}+{1\over p_{561}^2}\Big)\nonumber\\
&&~~~~~~~~~~+\Big({1\over \ell^2(-2\ell\cdot
p_{56}+p_{56}^2)}+{1\over \ell^2(-2\ell\cdot
p_{1234}+p_{1234}^2)}\Big)\Big({1\over p_{123}^2}+{1\over
p_{234}^2}\Big)~,~~~\nonumber\eea
as well as
\bea &&{1\over
\ell^2}(\cT^{\cQ}_{11}+\cT^{\cQ}_{15,2}+\cT^{\cQ}_{13}+\cT^{\cQ}_{15,1})+{1\over
\ell^2}(\cT^{\cQ}_{31}+\cT^{\cQ}_{35,2}+\cT^{\cQ}_{32}+\cT^{\cQ}_{35,1})\nonumber\\
&&~~~~~~~~~~~~={1\over \ell^2(-2\ell\cdot
p_{12}+p_{12}^2)}\Big({1\over p_{345}^2}+{1\over
p_{456}^2}\Big)+{1\over \ell^2(-2\ell\cdot
p_{3456}+p_{3456}^2)}\Big({1\over p_{345}^2}+{1\over
p_{456}^2}\Big)~,~~~\eea
and
\bea &&{1\over
\ell^2}(\cT^{\cQ}_{33}+\cT^{\cQ}_{36,2}+\cT^{\cQ}_{34}+\cT^{\cQ}_{36,1})+{1\over
\ell^2}(\cT^{\cQ}_{52}+\cT^{\cQ}_{55,2}+\cT^{\cQ}_{53}+\cT^{\cQ}_{55,1})\nonumber\\
&&~~~~~~~~~~~~={1\over \ell^2(-2\ell\cdot
p_{34}+p_{34}^2)}\Big({1\over p_{561}^2}+{1\over
p_{612}^2}\Big)+{1\over \ell^2(-2\ell\cdot
p_{5612}+p_{5612}^2)}\Big({1\over p_{561}^2}+{1\over
p_{612}^2}\Big)~,~~~\eea
and
\bea &&{1\over
\ell^2}(\cT^{\cQ}_{42}+\cT^{\cQ}_{45,2}+\cT^{\cQ}_{43}+\cT^{\cQ}_{45,1})+{1\over
\ell^2}(\cT^{\cQ}_{61}+\cT^{\cQ}_{66,2}+\cT^{\cQ}_{62}+\cT^{\cQ}_{66,1})\nonumber\\
&&~~~~~~~~~~~~={1\over \ell^2(-2\ell\cdot
p_{45}+p_{45}^2)}\Big({1\over p_{612}^2}+{1\over
p_{123}^2}\Big)+{1\over \ell^2(-2\ell\cdot
p_{6123}+p_{6123}^2)}\Big({1\over p_{612}^2}+{1\over
p_{123}^2}\Big)~,~~~\eea
and
\bea &&{1\over
\ell^2}(\cT^{\cQ}_{63}+\cT^{\cQ}_{65,2}+\cT^{\cQ}_{64}+\cT^{\cQ}_{65,1})+{1\over
\ell^2}(\cT^{\cQ}_{22}+\cT^{\cQ}_{25,2}+\cT^{\cQ}_{23}+\cT^{\cQ}_{25,1})\nonumber\\
&&~~~~~~~~~~~~={1\over \ell^2(-2\ell\cdot
p_{61}+p_{61}^2)}\Big({1\over p_{234}^2}+{1\over
p_{345}^2}\Big)+{1\over \ell^2(-2\ell\cdot
p_{2345}+p_{2345}^2)}\Big({1\over p_{234}^2}+{1\over
p_{345}^2}\Big)~.~~~\eea
Thus we confirm the equivalence among results of Feynman diagram
method, $\cQ$-cut representation and recursive formula
(\ref{recurStepFinal}) term by term. In fact, by cyclic invariance,
we can rewrite the integrand (\ref{FD6ptPhi4}) as
\bea \cI^{\cF}&=&\Big\{{1\over \ell^2(-2\ell\cdot
p_{12}+p_{12}^2)(-2\ell\cdot p_{1234}+p_{1234}^2)}+{1\over
\ell^2(-2\ell\cdot p_{12}+p_{12}^2)}\Big({1\over p_{345}^2}+{1\over
p_{456}^2}\Big)\nonumber\\
&&+{1\over \ell^2(-2\ell\cdot p_{1234}+p_{1234}^2)}\Big({1\over
p_{123}^2}+{1\over
p_{123}^2}\Big)\Big\}+\cyclic{1,2,3,4,5,6}~.~~~\eea
Then the recursive formula of tree diagram
$\cT^{\cQ}(\ell,-\ell,\sigma_1,\sigma_2,\sigma_3,\sigma_4,\sigma_5,\sigma_6)$
reproduces the result under the same ordering in $\cI^{\cF}$. For
instance, result of $\cT^{\cQ}(\ell,-\ell,1,2,3,4,5,6)$ reproduces
the above result in the curly bracket.

%%%%%%%%%%%%%%%%%%
\subsection{The one-loop four-point amplitude in scalar $\phi^3$ theory}
%%%%%%%%%%%%%%%%%%%

%%%%%%%%%%%%%%%%%%%%%%%%
\begin{figure}
  \centering
  \includegraphics[width=6.5in]{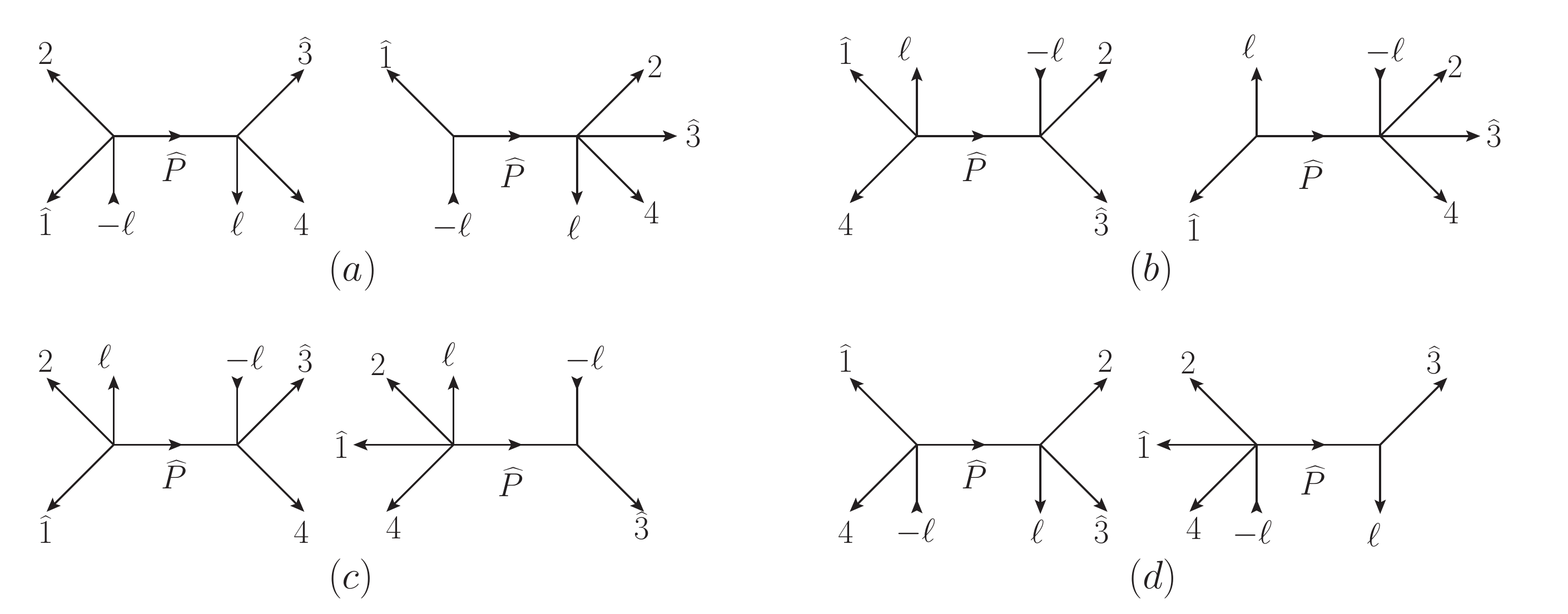}\\
  \caption{Non-vanishing diagrams for (a)$\cT_1(\ell,-\ell,1,2,3,4)$, (b)$\cT_2(\ell,-\ell,2,3,4,1)$,
  (c)$\cT_3(\ell,-\ell,3,4,1,2)$, (d)$\cT_4(\ell,-\ell,4,1,2,3)$, under $p_1,p_3$ BCFW deformation.}\label{FigfourPtPhi3BCFW}
\end{figure}
%%%%%%%%%%%%%%%%%%%%%%

Let us now discuss the integrand of one-loop four-point amplitude in color-ordered
scalar $\phi^3$ theory, so the tree diagram
$\cT^{\cQ}$ have four contributions, denoted as
\bea
\cT^{\cQ}=\cT_1^{\cQ}(\ell,-\ell,1,2,3,4)+\cT_2^{\cQ}(\ell,-\ell,2,3,4,1)
+\cT_3^{\cQ}(\ell,-\ell,3,4,1,2)+\cT_4^{\cQ}(\ell,-\ell,4,1,2,3)~.~~~\eea
The momentum deformation is taken as
$$\widehat{p}_1=p_1+zq~~,~~\widehat{p}_3=p_3-zq~~,~~q\cdot p_{1,3}=q^2=0~.~~$$
%Since $\cI^{\mathcal{Q}}_2=\cI^{\mathcal{Q}}_3=0$, the contributions
%$\cR^{\mathcal{Q}}_{A,1},\cR^{\mathcal{Q}}_{A,2}$ will not present.
%However, all $\cR'_{B,1}$, $\cR''_{B,1}$, $\cR'''_{B,1}$,
%$\cR'_{B,2}$, $\cR''_{B,2}$, $\cR'''_{B,2}$ will contribute. Since
%again we are considering color-ordered amplitude, the tree diagram
%$\cT^{\cQ}$ have four contributions, denoted as
%
%\bea
%\cT^{\cQ}=\cT_1^{\cQ}(\ell,-\ell,1,2,3,4)+\cT_2^{\cQ}(\ell,-\ell,2,3,4,1)+\cT_3^{\cQ}(\ell,-\ell,3,4,1,2)+\cT_4^{\cQ}(\ell,-\ell,4,1,2,3)~.~~~\eea
%
Under the given momentum deformation, each $\cT_{i}^{\cQ}$ has two
non-vanishing terms\footnote{Since the one-loop integrand $\cI^{\mathcal{Q}}_2=\cI^{\mathcal{Q}}_3=0$, the contributions to $\cR^{\mathcal{Q}}_{A,1},\cR^{\mathcal{Q}}_{A,2}$ will be zero.
However, all $\cR'_{B,1}$, $\cR''_{B,1}$, $\cR'''_{B,1}$, $\cR'_{B,2}$, $\cR''_{B,2}$, $\cR'''_{B,2}$ will contribute. }, as shown in Figure \ref{FigfourPtPhi3BCFW}.
Recall that the integrand of one-loop four-point amplitude in scalar
$\phi^3$ theory, after partial fraction identity, is given by
\cite{Huang:2015cwh}
\bea \cI^{\cF}(1,2,3,4)={1\over \ell^2}\Big({1\over -2\ell\cdot
p_1}+{1\over p_{12}^2}\Big){1\over -2\ell\cdot
p_{12}+p_{12}^2}\Big({1\over 2\ell\cdot p_4}+{1\over
p_{34}^2}\Big)+\cyclic{1,2,3,4}~.~~~\label{FD4ptPhi3}\eea
We want to show that, the integrand given by recursive formula
(\ref{recurStepFinal}) is equivalent to (\ref{FD4ptPhi3}), up to
certain scale free terms.

Let us start by computing the two diagrams in Figure
\ref{FigfourPtPhi3BCFW}.a. The four-point tree amplitude is
\bea \cA_4(1,2,3,4)={1\over (p_1+p_2)^2}+{1\over
(p_2+p_3)^2}~,~~~\eea
and let us define
\bea z_{12}^{\pm}\equiv -{\pm 2\ell\cdot p_{12}+p_{12}^2\over
2q\cdot(p_{12}\pm \ell)}~~,~~z_{41}^{\pm}\equiv -{\pm 2\ell\cdot
p_{41}+p_{41}^2\over 2q\cdot(p_{41}\pm \ell)}~~,~~z_1\equiv
-{2\ell\cdot p_1\over 2q\cdot\ell}~~,~~z_3\equiv {2\ell\cdot
p_3\over 2q\cdot\ell}~.~~~\eea
The first diagram gives a $\mathcal{R}'_{B,2}$-type contribution,
\bea \cT_{11}^{\cQ}&=&\cA_4(-\ell,\WH{1},2,\WH{P}){1\over
-2\ell\cdot
p_{12}+p_{12}^2}\cA_4(-\WH{P},\WH{3},4,\ell)\nonumber\\
&=&\Big({1\over -2\ell\cdot \WH{p}_1}+{1\over
\WH{p}_{12}^2}\Big){1\over -2\ell\cdot p_{12}+p_{12}^2}\Big({1\over
2\ell\cdot p_4}+{1\over \WH{p}_{34}^2}\Big)\Big|_{z_{12}^-}\equiv
\cT^{\cQ}_{11,1}+\cT^{\cQ}_{11,2}+\cT^{\cQ}_{11,3}+\cT^{\cQ}_{11,4}~,~~~\eea
where $z=z_{12}^-$, and $\mathcal{T}^{\mathcal{Q}}_{11,i}$ denotes
the four terms after expanding the result. The second diagram gives
a $\mathcal{R}''_{B,2}$-type contribution,
\bea \cT_{12}^{\cQ}&=&\cA_3(-\ell,\WH{1},\WH{P}){1\over -2\ell\cdot
p_1}\cA_3(-\WH{P},2,P'){1\over 2\ell\cdot
\WH{p}_{34}+\WH{p}_{34}^2}\cA_4(-P',\WH{3},4,\alpha\ell)\nonumber\\
&=&{1\over -2\ell\cdot p_1}{1\over -2\ell\cdot
\WH{p}_{12}+\WH{p}_{12}^2}\Big({1\over \alpha(2\ell\cdot
p_4)}+{1\over
\WH{p}_{34}^2}\Big)\Big|_{z=z_1,\alpha=\alpha_{12}}~,~~~\eea
where $\WH{P},P'$ are understood to follow the momentum conservation
of each sub-amplitudes, and
\bea \alpha_{12}=-{\WH{p}_{34}^2\over 2\ell\cdot
\WH{p}_{34}}={\WH{p}_{12}^2\over 2\ell\cdot
\WH{p}_{12}}\Big|_{z=z_1}~.~~~\eea
In fact, when substituting $\alpha_{12}$ back in $\cT_{12}^{\cQ}$,
we get
\bea \cT_{12}^{\cQ}&=&{1\over -2\ell\cdot p_1}{1\over -2\ell\cdot
\WH{p}_{12}+\WH{p}_{12}^2}\Big(-{-2\ell\cdot
\WH{p}_{12}+\WH{p}_{12}^2\over \WH{p}_{12}^2(2\ell\cdot
p_4)}+{1\over 2\ell\cdot p_4}+{1\over
\WH{p}_{34}^2}\Big)\Big|_{z=z_1}\nonumber\\
&=&-{1\over -2\ell\cdot p_1}{1\over \WH{p}_{12}^2(2\ell\cdot
p_4)}\Big|_{z=z_1}+{1\over -2\ell\cdot p_1}{1\over -2\ell\cdot
\WH{p}_{12}+\WH{p}_{12}^2}\Big({1\over 2\ell\cdot p_4}+{1\over
\WH{p}_{34}^2}\Big)\Big|_{z=z_1}~.~~~\label{fourPtPhi3T12}\eea
Note that
\bea \WH{p}_{12}^2|_{z_1}=p_{12}^2+z_1(2q\cdot p_{12})={2K_1\cdot
\ell\over 2q\cdot \ell}~~,~~K_1\equiv (p_{12}^2)q-(2q\cdot
p_{12})p_1~,~~~\eea
so the first term in (\ref{fourPtPhi3T12}) is a scale free term and
can be ignored. Hence we have four terms from $\cT_{11}^{\cQ}$ and
two terms from $\cT_{12}^{\cQ}$, and we want to compare the sum
${1\over
\ell^2}(\cT_{11,1}^{\cQ}+\cT_{11,2}^{\cQ}+\cT_{11,3}^{\cQ}+\cT_{11,4}^{\cQ}+\cT_{12,1}^{\cQ}+\cT_{12,2}^{\cQ})$
with
\bea {1\over \ell^2}\Big({1\over -2\ell\cdot p_1}+{1\over
p_{12}^2}\Big){1\over -2\ell\cdot p_{12}+p_{12}^2}\Big({1\over
2\ell\cdot p_4}+{1\over p_{34}^2}\Big)\equiv
\cI^{\cF}_{1,1}+\cI^{\cF}_{1,2}+\cI^{\cF}_{1,3}+\cI^{\cF}_{1,4}~.~~~\eea
To see the correspondence explicitly, firstly we have
\bea &&\cT^{\cQ}_{11,1}+\cT^{\cQ}_{12,1}\\
&=&{1\over -2\ell\cdot \WH{p}_1}{1\over -2\ell\cdot
p_{12}+p_{12}^2}{1\over 2\ell\cdot p_4}\Big|_{z_{12}^-}+{1\over
-2\ell\cdot p_1}{1\over -2\ell\cdot
\WH{p}_{12}+\WH{p}_{12}^2}{1\over 2\ell\cdot
p_4}\Big|_{z_{1}}\nonumber\\
&=&\Big({1\over (-2\ell\cdot p_1)+\lambda(-2\ell\cdot
p_{12}+p_{12}^2)}{1\over (-2\ell\cdot p_{12}+p_{12}^2)}+{1\over
(-2\ell\cdot p_1)}{1\over (-2\ell\cdot p_{12}+p_{12}^2)+(-2\ell\cdot
p_1)/\lambda}\Big){1\over 2\ell\cdot p_4}\nonumber\\
&=&{1\over -2\ell\cdot p_1}{1\over -2\ell\cdot
p_{12}+p_{12}^2}{1\over 2\ell\cdot p_4}~,~~~\nonumber\eea
where $\lambda={2q\cdot\ell\over 2q\cdot(p_{12}-\ell)}$, and in the
last line we have used the identity (\ref{idenAB}). So we see that
\bea {1\over
\ell^2}(\cT^{\cQ}_{11,1}+\cT^{\cQ}_{12,1})=\cI^{\cF}_{1,1}~.~~~\eea
Next, we have
\bea {1\over \ell^2}\cT^{\cQ}_{11,3}-\cI^{\cF}_{1,3}&=&{1\over
\ell^2}\Big({1\over \WH{p}_{12}^2}{1\over -2\ell\cdot
p_{12}+p_{12}^2}{1\over 2\ell\cdot p_4}\Big|_{z_{12}^-}-{1\over
p_{12}^2}{1\over -2\ell\cdot
p_{12}+p_{12}^2}{1\over 2\ell\cdot p_4}\Big)\nonumber\\
&=&-{2q\cdot p_{12}\over \ell^2(2K_{12}\cdot \ell)(2\ell\cdot
p_4)p_{12}^2}~~~,~~~K_{12}\equiv (p_{12}^2)q-(2q\cdot
p_{12})p_{12}~.~~~\eea
So ${1\over \ell^2}\cT_{11,3}^{\cQ}$ is equivalent to
$\cI^{\cF}_{1,3}$, up to a scale free term. Similarly,
\bea {1\over \ell^2}\cT^{\cQ}_{11,4}-\cI^{\cF}_{1,4}&=&{1\over
\ell^2}\Big({1\over \WH{p}_{12}^2}{1\over -2\ell\cdot
p_{12}+p_{12}^2}{1\over \WH{p}_{34}^2}\Big|_{z_{12}^-}-{1\over
p_{12}^2}{1\over -2\ell\cdot p_{12}+p_{12}^2}{1\over
p_{34}^2}\Big)\nonumber\\
&=&-{(2q\cdot p_{12})(2q\cdot \ell)\over \ell^2(2K_{12}\cdot
\ell)^2p_{12}^2}+{(2q\cdot p_{12})^2\over \ell^2(2K_{12}\cdot
\ell)^2p_{12}^2}-{2q\cdot p_{12}\over \ell^2(2K_{12}\cdot
\ell)(p_{12}^2)^2}~,~~~\eea
which is also a scale free term.

Finally, we have
\bea &&{1\over
\ell^2}(\cT_{11,2}^{\cQ}+\cT_{12,2}^{\cQ})-\cI^{\cF}_{1,2}\\
&=&{1\over \ell^2}\Big({1\over -2\ell\cdot \WH{p}_1}{1\over
-2\ell\cdot p_{12}+p_{12}^2}{1\over
\WH{p}_{34}^2}\Big|_{z_{12}^-}+{1\over -2\ell\cdot p_1}{1\over
-2\ell\cdot \WH{p}_{12}+\WH{p}_{12}^2}{1\over
\WH{p}_{34}^2}\Big|_{z_{1}}-{1\over -2\ell\cdot p_1}{1\over
-2\ell\cdot p_{12}+p_{12}^2}{1\over
p_{34}^2}\Big)\nonumber\\
&=&{1\over \ell^2}\Big({p_{12}^2(2q\cdot p_{12}-2q\cdot
\ell)^2(2\ell \cdot p_1)F_3+p_{12}^2(2q\cdot \ell)^2(-2\ell\cdot
p_{12}+p_{12}^2)F_2-F_1F_2F_3\over p_{12}^2(-2\ell\cdot
p_1)(-2\ell\cdot p_{12}+p_{12}^2)F_1F_2F_3}\Big)\nonumber\\
&=&-{(2q\cdot p_{12})^2\over \ell^2(2K_1\cdot \ell)(2K_{12}\cdot
\ell)p_{12}^2}~,~~~\nonumber\eea
where
\bea &&F_1\equiv p_{12}^2(2q\cdot \ell)-(2\ell\cdot p_1)(2q\cdot
p_{12})-(2\ell\cdot p_2)(2q\cdot \ell)~,~~~\\
&&F_2\equiv 2K_{12}\cdot \ell=p_{12}^2(2q\cdot \ell)-(2\ell\cdot
p_1)(2q\cdot p_{12})-(2\ell\cdot p_2)(2q\cdot p_{12})~,~~~\\
&&F_3\equiv 2K_1\cdot\ell=p_{12}^2(2q\cdot \ell)-(2\ell\cdot
p_1)(2q\cdot p_{12})~.~~~\eea
Thus we conclude that
\bea {1\over \ell^2}\cT^{\cQ}_{1}&=&\cI^{\cF}_1+{2q\cdot \ell\over
\ell^2(-2\ell\cdot p_1)(2K_1\cdot \ell)(2\ell\cdot p_4)}-{2q\cdot
p_{12}\over \ell^2(2K_{12}\cdot \ell)(2\ell\cdot
p_4)p_{12}^2}-{(2q\cdot p_{12})(2q\cdot \ell)\over
\ell^2(2K_{12}\cdot \ell)^2p_{12}^2}\nonumber\\
&&~~~~~~~~~~~~~~~~~~~~~~~+{(2q\cdot p_{12})^2\over
\ell^2(2K_{12}\cdot \ell)^2p_{12}^2}-{2q\cdot p_{12}\over
\ell^2(2K_{12}\cdot \ell)(p_{12}^2)^2}-{(2q\cdot p_{12})^2\over
\ell^2(2K_1\cdot \ell)(2K_{12}\cdot \ell)p_{12}^2}~.~~~\nonumber\eea
It confirms that, the result of recursive formula
(\ref{recurStepFinal}) is equivalent to the result of Feynman
diagram method, up to some scale free terms.

The same computation can be applied to tree diagrams $\cT_{2}^{\cQ},
\cT_{3}^{\cQ}$ and $\cT_{4}^{\cQ}$. For $\cT_{2}^{\cQ}$, we have two
contributing diagrams as shown in Figure \ref{FigfourPtPhi3BCFW}.b,
and we get
\bea \cT_{21}^{\cQ}&=&\cA_4(4,\WH{1},\ell,\WH{P}){1\over 2\ell\cdot
p_{41}+p_{41}^2}\cA_4(-\WH{P},-\ell,2,\WH{3})\nonumber\\
&=&\Big({1\over 2\ell \cdot \WH{p}_1}+{1\over
\WH{p}_{41}^2}\Big){1\over -2\ell\cdot p_{23}+p_{23}^2}\Big({1\over
-2\ell\cdot p_2}+{1\over \WH{p}_{23}^2}\Big)\Big|_{z_{41}^+}\equiv
\cT_{21,1}^{\cQ}+\cT_{21,2}^{\cQ}+\cT_{21,3}^{\cQ}+\cT_{21,4}^{\cQ}~,~~~\eea
as well as
\bea \cT_{22}^{\cQ}&=&\cA_3(\WH{1},\ell,\WH{P}){1\over 2\ell\cdot
p_1}\cA_4(-\alpha \ell,2,\WH{3},P'){1\over -2\ell\cdot
\WH{p}_{23}+\WH{p}_{23}^2}\cA_3(-P',4,-\WH{P})\nonumber\\
&=&{2 q\cdot \ell\over (2K_1'\cdot \ell)(2\ell\cdot p_1)(2\ell\cdot
p_2)}+{1\over 2\ell\cdot p_1}\Big({1\over -2\ell\cdot p_2}+{1\over
\WH{p}_{23}^2}\Big){1\over -2\ell \cdot
\WH{p}_{23}+\WH{p}_{23}^2}\Big|_{z_{1}}~,~~~\label{fourPtPhi3T22}\eea
where $K'_1\equiv (p_{23}^2)q+(2q\cdot p_{23})p_1$,
$$\alpha_{23}={\WH{p}_{23}^2\over 2\ell\cdot
\WH{p}_{23}}\Big|_{z=z_{1}}~.$$ The first term in
(\ref{fourPtPhi3T22}) is scale free, while the second and third
terms are denoted as $\cT_{22,1}^{\cQ},\cT_{22,2}^{\cQ}$. The result
${1\over \ell^2}\cT_{2}^{\cQ}$ is equivalent to
\bea {1\over \ell^2}\Big({1\over -2\ell\cdot p_2}+{1\over
p_{23}^2}\Big){1\over -2\ell\cdot p_{23}+p_{23}^2}\Big({1\over
2\ell\cdot p_1}+{1\over p_{41}^2}\Big)\equiv
\cI^{\cF}_{2,1}+\cI^{\cF}_{2,2}+\cI^{\cF}_{2,3}+\cI^{\cF}_{2,4}~,~~~\eea
up to some scale free terms. To see this, we have
\bea &&{1\over
\ell^2}(\cT_{21,1}^{\cQ}+\cT_{22,1}^{\cQ})=\cI_{2,1}^{\cF}~,~~~\\
&&{1\over \ell^2}\cT_{21,3}^{\cQ}=\cI_{2,2}^{\cF}+{2q\cdot
p_{23}\over \ell^2(2K_{23}\cdot \ell)(2\ell\cdot
p_2)p_{23}^2}~~,~~K_{23}\equiv (p_{23}^2)q-(2q\cdot
p_{23})p_{23}~,~~~\\
&&{1\over \ell^2}\cT_{21,4}^{\cQ}=\cI_{2,4}^{\cF}-{2q\cdot
p_{23}\over \ell^2(2K_{23}\cdot \ell)(p_{23}^2)^2}+{(2q\cdot
p_{23})^2\over \ell^2(2K_{23}\cdot \ell)^2p_{23}^2}-{(2q\cdot
p_{23})(2q\cdot \ell)\over \ell^2(2K_{23}\cdot
\ell)^2p_{23}^2}~,~~~\eea
and
\bea {1\over
\ell^2}(\cT_{21,2}^{\cQ}+\cT_{22,2}^{\cQ})=\cI_{2,3}^{\cF}-{(2q\cdot
p_{23})^2\over \ell^2(2K'_{1}\cdot \ell)(2K_{23}\cdot
\ell)p_{23}^2}~.~~~\eea
Thus confirming the equivalence.

For tree diagram $\cT_{3}^{\cQ}$, we have two contributing diagrams
as shown in Figure \ref{FigfourPtPhi3BCFW}.c, and we get
\bea \cT_{31}^{\cQ}&=&\cA_4(\WH{1},2,\ell,\WH{P}){1\over 2\ell\cdot
p_{12}+p_{12}^2}\cA_4(-\WH{P},-\ell,\WH{3},4)\nonumber\\
&=&\Big({1\over 2\ell \cdot p_2}+{1\over \WH{p}_{12}^2}\Big){1\over
-2\ell\cdot p_{34}+p_{34}^2}\Big({1\over -2\ell\cdot
\WH{p}_3}+{1\over \WH{p}_{34}^2}\Big)\Big|_{z_{12}^+}\equiv
\cT_{31,1}^{\cQ}+\cT_{31,2}^{\cQ}+\cT_{31,3}^{\cQ}+\cT_{31,4}^{\cQ}~,~~~\eea
as well as
\bea \cT_{32}^{\cQ}&=&\cA_3(\WH{P},4,P'){1\over 2\ell\cdot
\WH{p}_{12}+\WH{p}_{12}^2}\cA_4(-P',\WH{1},2,\alpha \ell){1\over
-2\ell\cdot
p_3}\cA_3(-\WH{P},-\ell,\WH{3})\nonumber\\
&=&{2q\cdot \ell\over (2\ell\cdot p_2)(2\ell\cdot p_3)(2K_3\cdot
\ell)}+{1\over -2\ell \cdot \WH{p}_{34}+\WH{p}_{34}^2}\Big({1\over
2\ell\cdot p_2}+{1\over \WH{p}_{12}^2}\Big){1\over -2\ell\cdot
p_3}\Big|_{z_{3}}~,~~~\label{fourPtPhi3T32}\eea
where $K_3\equiv (p_{34}^2)q-(2q\cdot p_{34})p_3$,
$$\alpha_{34}={\WH{p}_{34}^2\over 2\ell\cdot
\WH{p}_{34}}\Big|_{z=z_{3}}~.$$ Again the first term in
(\ref{fourPtPhi3T32}) is scale-free, while the second and third term
are denoted as $\cT_{32,1}^{\cQ},\cT_{32,2}^{\cQ}$. The result
${1\over \ell^2}\cT_{3}^{\cQ}$ is equivalent to
\bea {1\over \ell^2}\Big({1\over -2\ell\cdot p_3}+{1\over
p_{34}^2}\Big){1\over -2\ell\cdot p_{34}+p_{34}^2}\Big({1\over
2\ell\cdot p_2}+{1\over p_{12}^2}\Big)\equiv
\cI^{\cF}_{3,1}+\cI^{\cF}_{3,2}+\cI^{\cF}_{3,3}+\cI^{\cF}_{3,4}~,~~~\eea
up to some scale free terms, which can be confirmed by
\bea &&{1\over
\ell^2}(\cT_{31,1}^{\cQ}+\cT_{32,1}^{\cQ})=\cI_{3,1}^{\cF}~,~~~\\
&&{1\over \ell^2}\cT_{31,2}^{\cQ}=\cI_{3,3}^{\cF}-{2q\cdot
p_{34}\over \ell^2(2K_{34}\cdot \ell)(2\ell\cdot
p_2)p_{34}^2}~~,~~K_{34}\equiv (p_{34}^2)q-(2q\cdot
p_{34})p_{34}~,~~~\\
&&{1\over \ell^2}\cT_{31,4}^{\cQ}=\cI_{3,4}^{\cF}-{2q\cdot
p_{34}\over \ell^2(2K_{34}\cdot \ell)(p_{34}^2)^2}+{(2q\cdot
p_{34})^2\over \ell^2(2K_{34}\cdot \ell)^2p_{34}^2}-{(2q\cdot
p_{34})(2q\cdot \ell)\over \ell^2(2K_{34}\cdot
\ell)^2p_{34}^2}~,~~~\eea
and
\bea {1\over
\ell^2}(\cT_{31,3}^{\cQ}+\cT_{32,2}^{\cQ})=\cI_{3,2}^{\cF}-{(2q\cdot
p_{34})^2\over \ell^2(2K_{3}\cdot \ell)(2K_{34}\cdot
\ell)p_{34}^2}~.~~~\eea

For tree diagram $\cT_{4}^{\cQ}$, we have two contributing diagrams
as shown in Figure \ref{FigfourPtPhi3BCFW}.d, and we get
\bea \cT_{41}^{\cQ}&=&\cA_4(-\ell,4,\WH{1},\WH{P}){1\over
-2\ell\cdot
p_{41}+p_{41}^2}\cA_4(-\WH{P},2,\WH{3},\ell)\nonumber\\
&=&\Big({1\over -2\ell \cdot p_4}+{1\over \WH{p}_{41}^2}\Big){1\over
-2\ell\cdot p_{41}+p_{41}^2}\Big({1\over 2\ell\cdot
\WH{p}_3}+{1\over \WH{p}_{23}^2}\Big)\Big|_{z_{41}^-}\equiv
\cT_{41,1}^{\cQ}+\cT_{41,2}^{\cQ}+\cT_{41,3}^{\cQ}+\cT_{41,4}^{\cQ}~,~~~\eea
as well as
\bea \cT_{42}^{\cQ}&=&\cA_4(-\alpha \ell,4,\WH{1},P'){1\over
-2\ell\cdot \WH{p}_{41}+\WH{p}_{41}^2}\cA_3(-P',2,\WH{P}){1\over
2\ell\cdot
p_3}\cA_3(-\WH{P},\WH{3},\ell)\nonumber\\
&=&{2q\cdot \ell\over (2\ell\cdot p_4)(2\ell\cdot p_3)(2K'_3\cdot
\ell)}+\Big({1\over -2\ell\cdot p_4}+{1\over
\WH{p}_{41}^2}\Big){1\over -2\ell \cdot
\WH{p}_{41}+\WH{p}_{41}^2}{1\over 2\ell\cdot
p_3}\Big|_{z_{3}}~,~~~\label{fourPtPhi3T42}\eea
where $K'_3\equiv (p_{41}^2)q+(2q\cdot p_{41})p_3$,
$$\alpha_{34}={\WH{p}_{41}^2\over 2\ell\cdot
\WH{p}_{41}}\Big|_{z=z_{3}}~.$$ The first term in
(\ref{fourPtPhi3T42}) is scale free, while the second and third
terms are denoted as $\cT_{42,1}^{\cQ},\cT_{42,2}^{\cQ}$. The result
${1\over \ell^2}\cT_{4}^{\cQ}$ is equivalent to
\bea {1\over \ell^2}\Big({1\over -2\ell\cdot p_4}+{1\over
p_{41}^2}\Big){1\over -2\ell\cdot p_{41}+p_{41}^2}\Big({1\over
2\ell\cdot p_3}+{1\over p_{23}^2}\Big)\equiv
\cI^{\cF}_{4,1}+\cI^{\cF}_{4,2}+\cI^{\cF}_{4,3}+\cI^{\cF}_{4,4}~,~~~\eea
up to some scale free terms, which can be confirmed by
\bea &&{1\over
\ell^2}(\cT_{41,1}^{\cQ}+\cT_{42,1}^{\cQ})=\cI_{4,1}^{\cF}~,~~~\\
&&{1\over \ell^2}\cT_{41,2}^{\cQ}=\cI_{4,2}^{\cF}+{2q\cdot
p_{41}\over \ell^2(2K_{41}\cdot \ell)(2\ell\cdot
p_4)p_{41}^2}~~,~~K_{41}\equiv (p_{41}^2)q-(2q\cdot
p_{41})p_{41}~,~~~\\
&&{1\over \ell^2}\cT_{41,4}^{\cQ}=\cI_{4,4}^{\cF}-{2q\cdot
p_{41}\over \ell^2(2K_{41}\cdot \ell)(p_{41}^2)^2}+{(2q\cdot
p_{41})^2\over \ell^2(2K_{41}\cdot \ell)^2p_{41}^2}-{(2q\cdot
p_{41})(2q\cdot \ell)\over \ell^2(2K_{41}\cdot
\ell)^2p_{41}^2}~,~~~\eea
and
\bea {1\over
\ell^2}(\cT_{41,3}^{\cQ}+\cT_{42,2}^{\cQ})=\cI_{4,3}^{\cF}-{(2q\cdot
p_{41})^2\over \ell^2(2K'_{3}\cdot \ell)(2K_{41}\cdot
\ell)p_{41}^2}~.~~~\eea

The above detailed computations shows that, the result of recursive
formula (\ref{recurStepFinal}) is equivalent to the one of Feynman
diagram method up to some scale free terms.

%%%%%%%%%%%%%%%%
\subsection{The one-loop four-point amplitude in Yang-Mills theory}
%%%%%%%%%%%%%%%%%%%

Now let us take a quick glance on the well studied example, the
one-loop four-gluon all plus helicity amplitude
$A^{\oneloop}(1^+,2^+,3^+,4^+)$ in planar Yang-Mills theory. The
integrand of the original $\mathcal{Q}$-cut representation, after
dropping some scale free terms, takes \cite{Baadsgaard:2015twa}
\bea \mathcal{I}^{\mathcal{Q}}\sim {\spbb{1~2}\spbb{3~4}\over
\spaa{1~2}\spaa{3~4}}{(\mu^2-\ell^2)^2\over \ell^2(2\ell\cdot
p_1)(p_{12}^2-2\ell\cdot p_{12})(-2\ell\cdot
p_4)}+\cyclic{1,2,3,4}~.~~~ \eea

From the perspective of recursive formula (\ref{recurStepFinal}),
the tree diagram $\mathcal{T}^{\mathcal{Q}}$ is a sum over four tree
diagrams, denoted as
\bea
\cT^{\cQ}&=&\cT_1^{\cQ}(\ell,-\ell,1^+,2^+,3^+,4^+)+\cT_2^{\cQ}(\ell,-\ell,2^+,3^+,4^+,1^+)
\nonumber\\
&&~~~~~~~+\cT_3^{\cQ}(\ell,-\ell,3^+,4^+,1^+,2^+)+\cT_4^{\cQ}(\ell,-\ell,4^+,1^+,2^+,3^+)~.~~~\label{YMQcut}\eea
To compute $\mathcal{T}^{\mathcal{Q}}_{i}$'s, we should choose an
appropriate momentum deformation. Different momentum deformation
leads to different factorization of these tree amplitudes. Although
the final result will be the same, the intermediate terms will be
quite different. We can choose a deformation such that the
computation is as simple as possible. Furthermore, the four
$\mathcal{T}^{\mathcal{Q}}_i$'s are in fact independent, so each
$\mathcal{T}^{\mathcal{Q}}_i$ could have its own momentum
deformation, which makes the computation more flexible. In the
following computations, we will take advantage of  this freedom.

Let us now take
$\mathcal{T}^{\mathcal{Q}}_1(\ell,-\ell,1^+,2^+,3^+,4^+)$ as
example, and assume the internal loop to be massive scalar for
simplicity\footnote{It is massive in 4-dim, but null in higher
dimension.}. Let us choose the following momentum deformation,
\bea |\widehat{2}\rangle
=|2\rangle-z|3\rangle~~~,~~~|\widehat{3}]=|3]+z|2]~.~~~\eea
Since by definition the one-loop integrand $\mathcal{I}^{\mathcal{Q}}_3=0$, we get only one
$\mathcal{R}'_{B,2}$-type contribution,
\bea
\mathcal{T}^{\mathcal{Q}}_1=\sum_{h}A(-\ell^{s},1^+,\widehat{2}^+,\widehat{P}^{h}){1\over
p_{12}^2-2\ell\cdot
p_{12}}A(-\widehat{P}^{-h},\widehat{3}^+,4^+,\ell^{s})~,~~~\eea
where the helicity sum is over all possible states $(+,-,s)$. From
the results of tree-level amplitudes in \cite{Huang:2015cwh}, we get
the non-vanishing contribution
\bea &&A(-\ell^{s},1^+,\widehat{2}^+,\widehat{P}^{s}){1\over
p_{12}^2-2\ell\cdot
p_{12}}A(-\widehat{P}^{s},\widehat{3}^+,4^+,\ell^{s})={\spbb{2~1}\over
\spaa{1~\widehat{2}}}{\mu^2-\ell^2\over \spab{1|-\ell|1}}{1\over
p_{12}^2-2\ell\cdot p_{12}}{\spbb{4~\widehat{3}}\over
\spaa{3~4}}{\mu^2-\ell^2\over \spab{4|\ell|4}}~,~~~ \eea
where $z={p_{12}^2-2\ell\cdot p_{12}\over 2q\cdot (p_{12}-\ell)}$.
Using the momentum conservation identity
\bea
p_1+\widehat{p}_2=-(\widehat{p}_3+p_4)~\to~(p_1+\widehat{p}_2)^2=(\widehat{p}_3+p_4)^2~\to~{\spbb{4~\widehat{3}}\over
\spaa{1~\widehat{2}}}={\spbb{2~1}\over \spaa{3~4}}={\spbb{4~3}\over
\spaa{1~2}}~,~~~ \eea
we instantly get
\bea \mathcal{T}^{\mathcal{Q}}_1={\spbb{1~2}\spbb{3,4}\over
\spaa{1~2}\spaa{3~4}}{(\mu^2-\ell^2)^2\over (-2\ell\cdot
p_1)(p^2_{12}-2\ell\cdot p_{12})(2\ell\cdot p_4) }~.~~~\eea
So ${1\over \ell^2}\mathcal{T}^{\mathcal{Q}}_{1}$ equals to a term
in (\ref{YMQcut}). Similarly, the BCFW deformation
$$\mathcal{T}^{\mathcal{Q}}_2(\ell,-\ell,2^+,\widehat{3}^+,\widehat{4}^+,1^+)~~,~~
\mathcal{T}^{\mathcal{Q}}_3(\ell,-\ell,3^+,\widehat{4}^+,\widehat{1}^+,2^+)~~,~~
\mathcal{T}^{\mathcal{Q}}_4(\ell,-\ell,4^+,\widehat{1}^+,\widehat{2}^+,3^+)$$
will produce the other three terms respectively. This simple example
is illustrative to show how the terms computed by recursive formula
(\ref{recurStepFinal}) are corresponding to the terms computed by
the original $\mathcal{Q}$-cut representation.

%%%%%%%%%%%%%%%%%%%%
\section{Conclusion}
\label{secConclusion}
%%%%%%%%%%%%%%%%%%%%%

In this note, we have taken initial steps for constructing one-loop integrand by combining
the BCFW deformation and the $\mathcal{Q}$-cut construction. We have obtained
a recursive formula (\ref{recurStepFinal}), where the one-loop integrand is
given by one-loop integrands with lower number of external
legs, and tree-level amplitudes. We have presented explicit examples to show the equivalence of our result
with the one given by Feynman diagrams and $\mathcal{Q}$-cut representation,
up to scale free terms.

%We found that, after getting rid of the forward singularity with dimensional deformation of loop momentum $\ell$, the BCFW deformation can be as well applied to reconstruct the one-loop integrand, resulting to a recursive formula (\ref{recurStepFinal}). It gives different factorization of one-loop integrand compared to scale deformation in the original $\mathcal{Q}$-cut construction, and the result is equivalent to the $\mathcal{Q}$-cut representation, possibly up to some scale free terms. In short, the BCFW deformation factorizes an one-loop integrand as products of tree amplitude and lower-point one-loop integrand, as well as products of two lower-point tree amplitudes. For the special terms where a sub-tree amplitude is three-point amplitude, the scale deformation is again required in order to exclude the contribution of massless bubbles. In these special cases, the terms are then factorized as products of three lower-point tree amplitudes. We confirm the recursive formula by presenting three examples, and comparing their result term by term with those of Feynman diagram method or the original $\mathcal{Q}$-cut representation.

There are several  possible applications of the recursive formula
(\ref{recurStepFinal}). The first one is to consider  the one-loop
factorization limit
$A_L^{\tree}A_R^{\oneloop}+A_L^{\oneloop}A_R^{\tree}+A_L^{\tree}\mathcal{S}A_R^{\tree}$.
It is easy to see that, in the recursive formula,
$\mathcal{R}^{\mathcal{Q}}_{A}$ contributes to the first two
factorization limits, while $\mathcal{R}^{\mathcal{Q}}_{B}$
contributes to the third term. The $\mathcal{R}^{\mathcal{Q}}_{B}$
part contains six terms, so naively the kernel $\mathcal{S}$ could
be very complicated. However, it could be the case that some terms
do not contribute, or their contributions simplify a lot in
the factorization limit. It would be interesting to investigate if
we can find some compact form for $\mathcal{S}$ or not. Using the
recursive formula, we can also study the behavior of integrands in certain
limits, for instance the single/double soft limit and the one-loop split
function. It is also possible to study the rational part of one-loop amplitudes when constructed using 4-dimensional unitarity cut method, especially if we
could write down some recursive relation for the rational part, based on our  formula. Finally, generalizations to higher loops and massive
external legs, which are a very important open questions in the original
$\mathcal{Q}$-cut representation, deserves to be investigated along this direction as well.

\section*{Acknowledgments}

BF, RH and ML is supported by Qiu-Shi Funding and the National
Natural Science Foundation of China (NSFC) with Grant No.11135006,
No.11125523 and No.11575156. RH would also like to acknowledge the
supporting from Chinese Postdoctoral Administrative Committee. SH
acknowledges support from the Thousand Young Talents program and the
Key Research Program of Frontier Sciences of CAS (Grant No.
QYZDBSSW-SYS014).

\appendix

%%%%%%%%%%%%%%%%%%%%%%%%%%%%%%%%%%%% %%%%%%%

\bibliographystyle{JHEP}
\bibliography{Qrecur}

\providecommand{\href}[2]{#2}\begingroup\raggedright\begin{thebibliography}{10}

\bibitem{Baadsgaard:2015twa}
C.~Baadsgaard, N.~E.~J. Bjerrum-Bohr, J.~L. Bourjaily, S.~Caron-Huot, P.~H.
  Damgaard, and B.~Feng, {\it {New Representations of the Perturbative
  S-Matrix}},  {\em Phys. Rev. Lett.} {\bf 116} (2016), no.~6 061601,
  [\href{http://arxiv.org/abs/1509.02169}{{\tt arXiv:1509.02169}}].

\bibitem{Huang:2015cwh}
R.~Huang, Q.~Jin, J.~Rao, K.~Zhou, and B.~Feng, {\it {The Q-cut Representation
  of One-loop Integrands and Unitarity Cut Method}},  {\em JHEP} {\bf 03}
  (2016) 057, [\href{http://arxiv.org/abs/1512.02860}{{\tt arXiv:1512.02860}}].

\bibitem{Geyer:2015bja}
Y.~Geyer, L.~Mason, R.~Monteiro, and P.~Tourkine, {\it {Loop Integrands for
  Scattering Amplitudes from the Riemann Sphere}},  {\em Phys. Rev. Lett.} {\bf
  115} (2015), no.~12 121603, [\href{http://arxiv.org/abs/1507.00321}{{\tt
  arXiv:1507.00321}}].

\bibitem{Geyer:2015jch}
Y.~Geyer, L.~Mason, R.~Monteiro, and P.~Tourkine, {\it {One-loop amplitudes on
  the Riemann sphere}},  {\em JHEP} {\bf 03} (2016) 114,
  [\href{http://arxiv.org/abs/1511.06315}{{\tt arXiv:1511.06315}}].

\bibitem{Baadsgaard:2015hia}
C.~Baadsgaard, N.~E.~J. Bjerrum-Bohr, J.~L. Bourjaily, P.~H. Damgaard, and
  B.~Feng, {\it {Integration Rules for Loop Scattering Equations}},  {\em JHEP}
  {\bf 11} (2015) 080, [\href{http://arxiv.org/abs/1508.03627}{{\tt
  arXiv:1508.03627}}].

\bibitem{He:2015yua}
S.~He and E.~Y. Yuan, {\it {One-loop Scattering Equations and Amplitudes from
  Forward Limit}},  {\em Phys. Rev.} {\bf D92} (2015), no.~10 105004,
  [\href{http://arxiv.org/abs/1508.06027}{{\tt arXiv:1508.06027}}].

\bibitem{Cachazo:2013gna}
F.~Cachazo, S.~He, and E.~Y. Yuan, {\it {Scattering equations and
  Kawai-Lewellen-Tye orthogonality}},  {\em Phys. Rev.} {\bf D90} (2014), no.~6
  065001, [\href{http://arxiv.org/abs/1306.6575}{{\tt arXiv:1306.6575}}].

\bibitem{Cachazo:2013hca}
F.~Cachazo, S.~He, and E.~Y. Yuan, {\it {Scattering of Massless Particles in
  Arbitrary Dimensions}},  {\em Phys. Rev. Lett.} {\bf 113} (2014), no.~17
  171601, [\href{http://arxiv.org/abs/1307.2199}{{\tt arXiv:1307.2199}}].

\bibitem{Cachazo:2013iea}
F.~Cachazo, S.~He, and E.~Y. Yuan, {\it {Scattering of Massless Particles:
  Scalars, Gluons and Gravitons}},  {\em JHEP} {\bf 07} (2014) 033,
  [\href{http://arxiv.org/abs/1309.0885}{{\tt arXiv:1309.0885}}].

\bibitem{Cachazo:2014nsa}
F.~Cachazo, S.~He, and E.~Y. Yuan, {\it {Einstein-Yang-Mills Scattering
  Amplitudes From Scattering Equations}},  {\em JHEP} {\bf 01} (2015) 121,
  [\href{http://arxiv.org/abs/1409.8256}{{\tt arXiv:1409.8256}}].

\bibitem{Cachazo:2014xea}
F.~Cachazo, S.~He, and E.~Y. Yuan, {\it {Scattering Equations and Matrices:
  From Einstein To Yang-Mills, DBI and NLSM}},  {\em JHEP} {\bf 07} (2015) 149,
  [\href{http://arxiv.org/abs/1412.3479}{{\tt arXiv:1412.3479}}].

\bibitem{Geyer:2016wjx}
Y.~Geyer, L.~Mason, R.~Monteiro, and P.~Tourkine, {\it {Two-Loop Scattering
  Amplitudes from the Riemann Sphere}},
  \href{http://arxiv.org/abs/1607.08887}{{\tt arXiv:1607.08887}}.

\bibitem{Cardona:2016bpi}
C.~Cardona and H.~Gomez, {\it {Elliptic scattering equations}},  {\em JHEP}
  {\bf 06} (2016) 094, [\href{http://arxiv.org/abs/1605.01446}{{\tt
  arXiv:1605.01446}}].

\bibitem{Gomez:2016bmv}
H.~Gomez, {\it {$\Lambda$ scattering equations}},  {\em JHEP} {\bf 06} (2016)
  101, [\href{http://arxiv.org/abs/1604.05373}{{\tt arXiv:1604.05373}}].

\bibitem{Cachazo:2015aol}
F.~Cachazo, S.~He, and E.~Y. Yuan, {\it {One-Loop Corrections from Higher
  Dimensional Tree Amplitudes}},  {\em JHEP} {\bf 08} (2016) 008,
  [\href{http://arxiv.org/abs/1512.05001}{{\tt arXiv:1512.05001}}].

\bibitem{Feng:2016nrf}
B.~Feng, {\it {CHY-construction of Planar Loop Integrands of Cubic Scalar
  Theory}},  {\em JHEP} {\bf 05} (2016) 061,
  [\href{http://arxiv.org/abs/1601.05864}{{\tt arXiv:1601.05864}}].

\bibitem{Britto:2004ap}
R.~Britto, F.~Cachazo, and B.~Feng, {\it {New recursion relations for tree
  amplitudes of gluons}},  {\em Nucl. Phys.} {\bf B715} (2005) 499--522,
  [\href{http://arxiv.org/abs/hep-th/0412308}{{\tt hep-th/0412308}}].

\bibitem{Britto:2005fq}
R.~Britto, F.~Cachazo, B.~Feng, and E.~Witten, {\it {Direct proof of tree-level
  recursion relation in Yang-Mills theory}},  {\em Phys. Rev. Lett.} {\bf 94}
  (2005) 181602, [\href{http://arxiv.org/abs/hep-th/0501052}{{\tt
  hep-th/0501052}}].

\bibitem{CaronHuot:2010zt}
S.~Caron-Huot, {\it {Loops and trees}},  {\em JHEP} {\bf 05} (2011) 080,
  [\href{http://arxiv.org/abs/1007.3224}{{\tt arXiv:1007.3224}}].

\bibitem{ArkaniHamed:2010kv}
N.~Arkani-Hamed, J.~L. Bourjaily, F.~Cachazo, S.~Caron-Huot, and J.~Trnka, {\it
  {The All-Loop Integrand For Scattering Amplitudes in Planar N=4 SYM}},  {\em
  JHEP} {\bf 01} (2011) 041, [\href{http://arxiv.org/abs/1008.2958}{{\tt
  arXiv:1008.2958}}].

\bibitem{Boels:2010nw}
R.~H. Boels, {\it {On BCFW shifts of integrands and integrals}},  {\em JHEP}
  {\bf 11} (2010) 113, [\href{http://arxiv.org/abs/1008.3101}{{\tt
  arXiv:1008.3101}}].

\bibitem{ArkaniHamed:2008yf}
N.~Arkani-Hamed and J.~Kaplan, {\it {On Tree Amplitudes in Gauge Theory and
  Gravity}},  {\em JHEP} {\bf 04} (2008) 076,
  [\href{http://arxiv.org/abs/0801.2385}{{\tt arXiv:0801.2385}}].

\bibitem{Cheung:2008dn}
C.~Cheung, {\it {On-Shell Recursion Relations for Generic Theories}},  {\em
  JHEP} {\bf 03} (2010) 098, [\href{http://arxiv.org/abs/0808.0504}{{\tt
  arXiv:0808.0504}}].

\bibitem{Feng:2009ei}
B.~Feng, J.~Wang, Y.~Wang, and Z.~Zhang, {\it {BCFW Recursion Relation with
  Nonzero Boundary Contribution}},  {\em JHEP} {\bf 01} (2010) 019,
  [\href{http://arxiv.org/abs/0911.0301}{{\tt arXiv:0911.0301}}].

\bibitem{Jin:2014qya}
Q.~Jin and B.~Feng, {\it {Recursion Relation for Boundary Contribution}},  {\em
  JHEP} {\bf 06} (2015) 018, [\href{http://arxiv.org/abs/1412.8170}{{\tt
  arXiv:1412.8170}}].

\bibitem{Jin:2015pua}
Q.~Jin and B.~Feng, {\it {Boundary Operators of BCFW Recursion Relation}},
  {\em JHEP} {\bf 04} (2016) 123, [\href{http://arxiv.org/abs/1507.00463}{{\tt
  arXiv:1507.00463}}].

\bibitem{Feng:2014pia}
B.~Feng, K.~Zhou, C.~Qiao, and J.~Rao, {\it {Determination of Boundary
  Contributions in Recursion Relation}},  {\em JHEP} {\bf 03} (2015) 023,
  [\href{http://arxiv.org/abs/1411.0452}{{\tt arXiv:1411.0452}}].

\bibitem{Feng:2015qna}
B.~Feng, J.~Rao, and K.~Zhou, {\it {On Multi-step BCFW Recursion Relations}},
  {\em JHEP} {\bf 07} (2015) 058, [\href{http://arxiv.org/abs/1504.06306}{{\tt
  arXiv:1504.06306}}].

\bibitem{Cheung:2015cba}
C.~Cheung, C.-H. Shen, and J.~Trnka, {\it {Simple Recursion Relations for
  General Field Theories}},  {\em JHEP} {\bf 06} (2015) 118,
  [\href{http://arxiv.org/abs/1502.05057}{{\tt arXiv:1502.05057}}].

\bibitem{Cheung:2015ota}
C.~Cheung, K.~Kampf, J.~Novotny, C.-H. Shen, and J.~Trnka, {\it {On-Shell
  Recursion Relations for Effective Field Theories}},  {\em Phys. Rev. Lett.}
  {\bf 116} (2016), no.~4 041601, [\href{http://arxiv.org/abs/1509.03309}{{\tt
  arXiv:1509.03309}}].

\end{thebibliography}\endgroup

\end{document}